\documentclass[letterpaper]{article} % DO NOT CHANGE THIS
\usepackage[preprint]{aaai2027}  % DO NOT CHANGE THIS
\usepackage[hyphens]{url}  % DO NOT CHANGE THIS
\usepackage{graphicx} % DO NOT CHANGE THIS
\usepackage{natbib}  % DO NOT CHANGE THIS AND DO NOT ADD ANY OPTIONS TO IT
\usepackage{caption} % DO NOT CHANGE THIS AND DO NOT ADD ANY OPTIONS TO IT
\usepackage{algorithm}
\usepackage{algorithmic}
\usepackage{multirow}
\usepackage{latexsym}
\usepackage{amsmath}
\usepackage{amssymb}
\usepackage[table]{xcolor}
\usepackage{booktabs}

\usepackage[most]{tcolorbox}
\tcbuselibrary{listings}
\usepackage{stmaryrd}
\usepackage[most]{tcolorbox}
\usepackage{booktabs}
\usepackage{xcolor}
\usepackage[T1]{fontenc}
\definecolor{slate}{RGB}{240, 242, 245}
\definecolor{groupgray}{RGB}{226,226,226}
\definecolor{blue1}{rgb}{0.8, 0.9, 1.0}
\definecolor{c1}{RGB}{132,167,64}
\definecolor{c12}{RGB}{109, 173, 209}
\definecolor{c8}{RGB}{36, 102, 248}
\definecolor{c3}{RGB}{238,155,000}
\definecolor{c4}{RGB}{183, 63, 66}
\definecolor{c5}{RGB}{0, 140, 140}
\definecolor{c6}{RGB}{251, 132, 002}
\definecolor{b1}{RGB}{252, 108, 133}
\definecolor{c7}{RGB}{248, 244, 237}
\definecolor{c2}{RGB}{68, 188, 249}
\definecolor{softblue}{rgb}{0.8, 0.9, 1.0}
\definecolor{softgreen}{rgb}{0.0, 0.6, 0.2}
\usepackage{newfloat}
\usepackage{listings}
\DeclareCaptionStyle{ruled}{labelfont=normalfont,labelsep=colon,strut=off} % DO NOT CHANGE THIS
\floatstyle{ruled}
\newfloat{listing}{tb}{lst}{}
\floatname{listing}{Listing}

\usepackage{booktabs}

\title{VIG-RL: Learning to Search and Insert for Verified Image Grounding}
\author {
    Qinhan Yu\textsuperscript{\rm 1}\equalcontrib,
    Jun Guang\textsuperscript{\rm 1}\equalcontrib,
    Chong Chen\textsuperscript{\rm 2}\corresponding,
    Wentao Zhang\textsuperscript{\rm 1}\corresponding
}
\affiliations {
    \textsuperscript{\rm 1}Peking University, \textsuperscript{\rm 2}Huawei Cloud BU\\
    yuqinhan@stu.pku.edu.cn, guangjun000000@gmail.com\\chenchong55@huawei.com, wentao.zhang@pku.edu.cn
}
\begin{document}

\maketitle

\begin{abstract}
In knowledge-intensive scenarios, providing reliable interleaved text-image responses requires Verified Image Grounding (VIG)—the precise integration of retrieved authentic visual evidence. Existing retrieval-augmented frameworks predominantly rely on decoupled, static pipelines, inherently failing to dynamically reason about \textit{when} external knowledge is required and \textit{where} visual assets should be contextually inserted. To bridge this gap, we propose \textbf{VIG-RL}, an autonomous agentic framework that formulates the search-selection-insertion workflow as an active decision-making process. Operating within a dynamic ReAct-style loop, VIG-RL is optimized via reinforcement learning, guided by a composite reward system that holistically evaluates the agent's step-by-step tool execution and final multimodal alignment. Extensive evaluations demonstrate that VIG-RL establishes a new state-of-the-art, significantly outperforming existing static baselines.

\end{abstract}

\section{Introduction}

\begin{figure}[t]
\centering
\includegraphics[width=\linewidth]{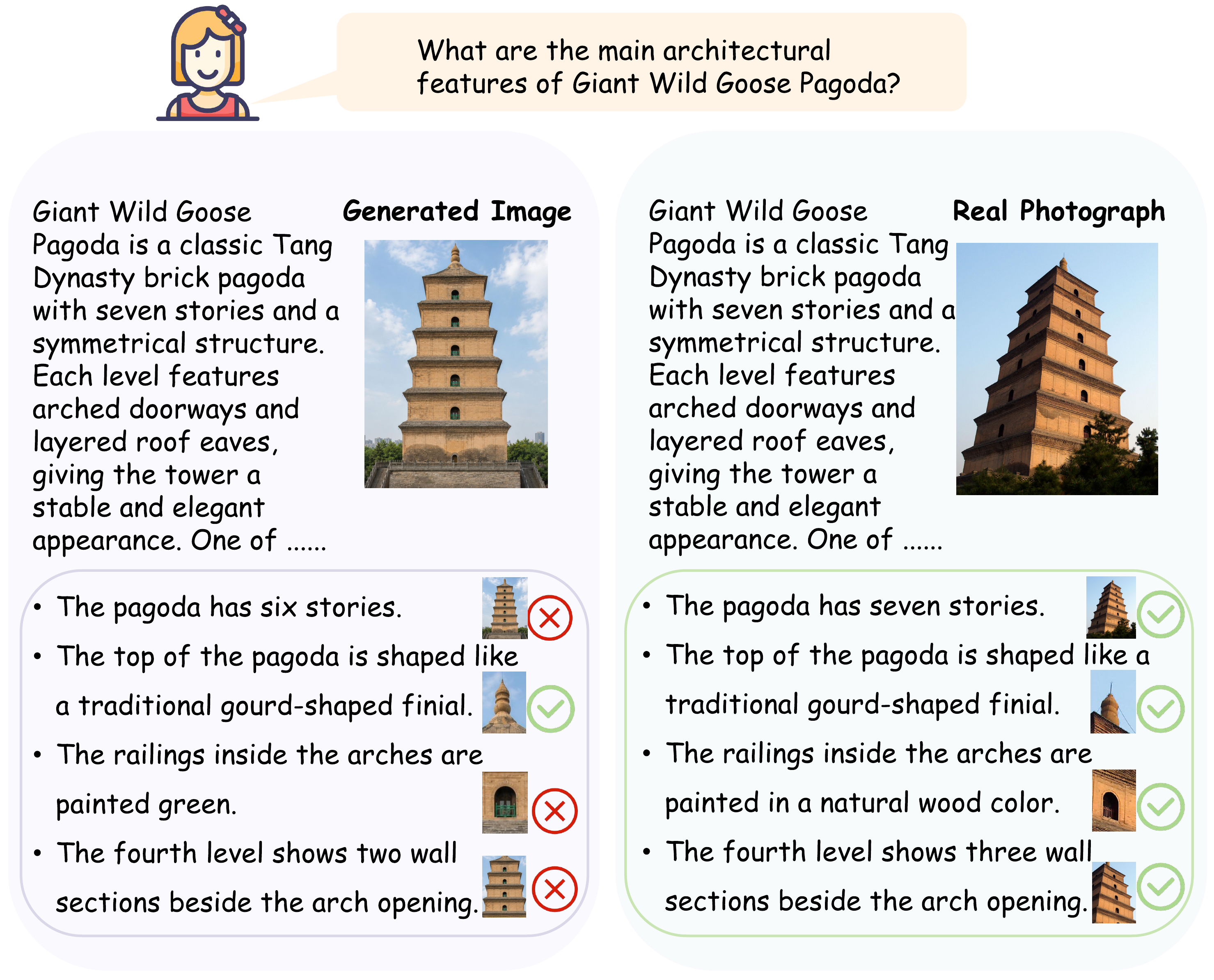}

\caption{\textbf{Verified Image Grounding (VIG) vs. Pure Generation.} The model-generated image of the Giant Wild Goose Pagoda contains architectural inconsistencies, including an incorrect number of stories, wrong railing colors, and missing structural details near the arch openings. In contrast, VIG retrieves authentic photographs, yielding factually grounded and visually consistent multimodal responses.}
\label{fig:examples}
\end{figure}

Large Language Models (LLMs)~\citep{brown2020language,touvron2023llama,liu2024deepseek} have demonstrated unparalleled proficiency in natural language understanding and generation. However, while contemporary Multimodal Large Language Models (MLLMs)~\citep{yin2024survey,hurst2024gpt,bai2025qwen3,comanici2025gemini} have achieved significant progress in processing multimodal inputs, their generation capabilities remain primarily restricted to single-modal text outputs. This limitation creates a critical gap in real-world scenarios, where users increasingly demand richer, reliably interleaved text-image responses to facilitate a more intuitive and comprehensive understanding of complex information.

To address these demands, recent advancements in interleaved multimodal generation~\citep{dong2024dreamllm,wu2024next,wang2024emu3,chen2025janus} rely predominantly on generative pipelines, whether through unified generative models or the invocation of external image generation tools, to synthesize visual content. While these approaches are capable of producing aesthetically coherent images, their outputs are inherently synthetic, prioritizing visual plausibility over objective factual accuracy. However, in knowledge-intensive scenarios, users explicitly demand factually-grounded, verified visual evidence rather than synthetic approximations. As illustrated in Figure~\ref{fig:examples}, when queried about the architectural features of the Giant Wild Goose Pagoda, a renowned historical landmark in China, a generative baseline renders a superficially plausible structure but catastrophically fails at a granular level, hallucinating an incorrect number of structural tiers, misrepresenting the coloration of the balustrades, and distorting the number of facets surrounding the arched openings. Such deceptive hallucinations corrupt the visual fidelity required for rigorous inquiry and mislead textual reasoning. Photorealistic synthetic images cannot substitute for authentic, verifiable real-world documentation retrieved dynamically from reliable archives.

This limitation of purely generative pipelines highlights the necessity of \textbf{Verified Image Grounding (VIG)}, where the core objective is to retrieve and seamlessly integrate contextually-aligned, authentic visual evidence into the generated text stream. In this context, "verified images" serve as reliable visual references, generally spanning from naturally captured photographs (e.g., historical landmarks) to human-vetted schematics (e.g., model architectures in academic papers). While recent retrieval-augmented frameworks~\citep{ma2024multi,zhu2025murar,yu2025mramg,xiao2025m2io} have attempted to incorporate such external visual evidence, they typically adopt a standard retrieve-then-generate paradigm. This decoupled approach makes it challenging to dynamically orchestrate fine-grained interactions during the generation process. To fulfill the stringent requirements of high-fidelity, interleaved multimodal generation, it is essential to develop dynamic reasoning capabilities that can precisely determine when to search, how to query, which evidence to select, and where to seamlessly insert it.

To bridge this gap, we propose VIG-RL, an agentic reinforcement learning framework tailored for the VIG task. Instead of relying on a decoupled retrieve-then-generate pipeline, VIG-RL trains an autonomous agent to dynamically orchestrate the entire search-and-insert process. Operating within a ReAct-style~\citep{yao2022react} interaction loop, the agent is equipped with a versatile action space: \textit{(i) text search} to acquire missing contextual knowledge, \textit{(ii) image search} to retrieve factual visual evidence, and \textit{(iii) answer generation} to compose the interleaved response. To effectively teach the agent exactly when to execute these actions and where to embed the retrieved visual identifiers, we optimize the agent via the GRPO algorithm~\citep{shao2024deepseekmath}. Driven by a multi-dimensional composite reward system that jointly evaluates precise format adherence, efficient search behavior, and the factual accuracy of the final answer, Reinforcement Learning (RL) efficiently aligns the agent's step-by-step reasoning. By explicitly learning to autonomously manage the provenance and integration of visual evidence, VIG-RL transcends static heuristics, charting a viable path for reliable, knowledge-intensive interleaved multimodal generation.
More precisely, our main contributions are summarized as follows:
\begin{itemize}
\item
We propose VIG-RL, an agentic framework for the VIG task. By shifting from static heuristics to a dynamic ReAct-style loop, it empowers the agent to autonomously orchestrate complex search and insertion actions.
\item
We design a specialized RL paradigm for multimodal agents using GRPO. Guided by a multi-dimensional composite reward system, this approach effectively aligns the agent's step-by-step reasoning to master precise retrieval and insertion policies.
\item
Extensive evaluations demonstrate that VIG-RL establishes a new state-of-the-art in interleaved multimodal generation under the VIG paradigm, consistently outperforming static retrieval-augmented baselines.
\end{itemize}

\section{Related Work}

\paragraph{Interleaved Multimodal Generation.}
Interleaved text--image generation has been explored through unified models~\citep{chern2024anole,wang2024emu3,xie2025show,zhou2025transfusion,chen2025janus} and tool-use frameworks~\citep{koh2023generating,dong2024dreamllm,wu2024next,shi2026duogen}, but remains prone to factual distortions in knowledge-intensive settings, motivating Verified Image Grounding (VIG) with retrieved authentic images. Recent retrieval-augmented methods have begun addressing VIG~\citep{ma2024multi,zhu2025murar}; MRAMG-Bench~\citep{yu2025mramg} is currently the only open-source benchmark with explicit ground-truth placement annotations, while M2IO-R1~\citep{xiao2025m2io} adopts a decoupled text-then-insertion pipeline. Unlike these static workflows, VIG-RL formulates VIG as an agentic process that jointly learns when to search, which evidence to select, and where to insert it.

\begin{figure*}[t]
    \centering    \includegraphics[width=0.95\linewidth]{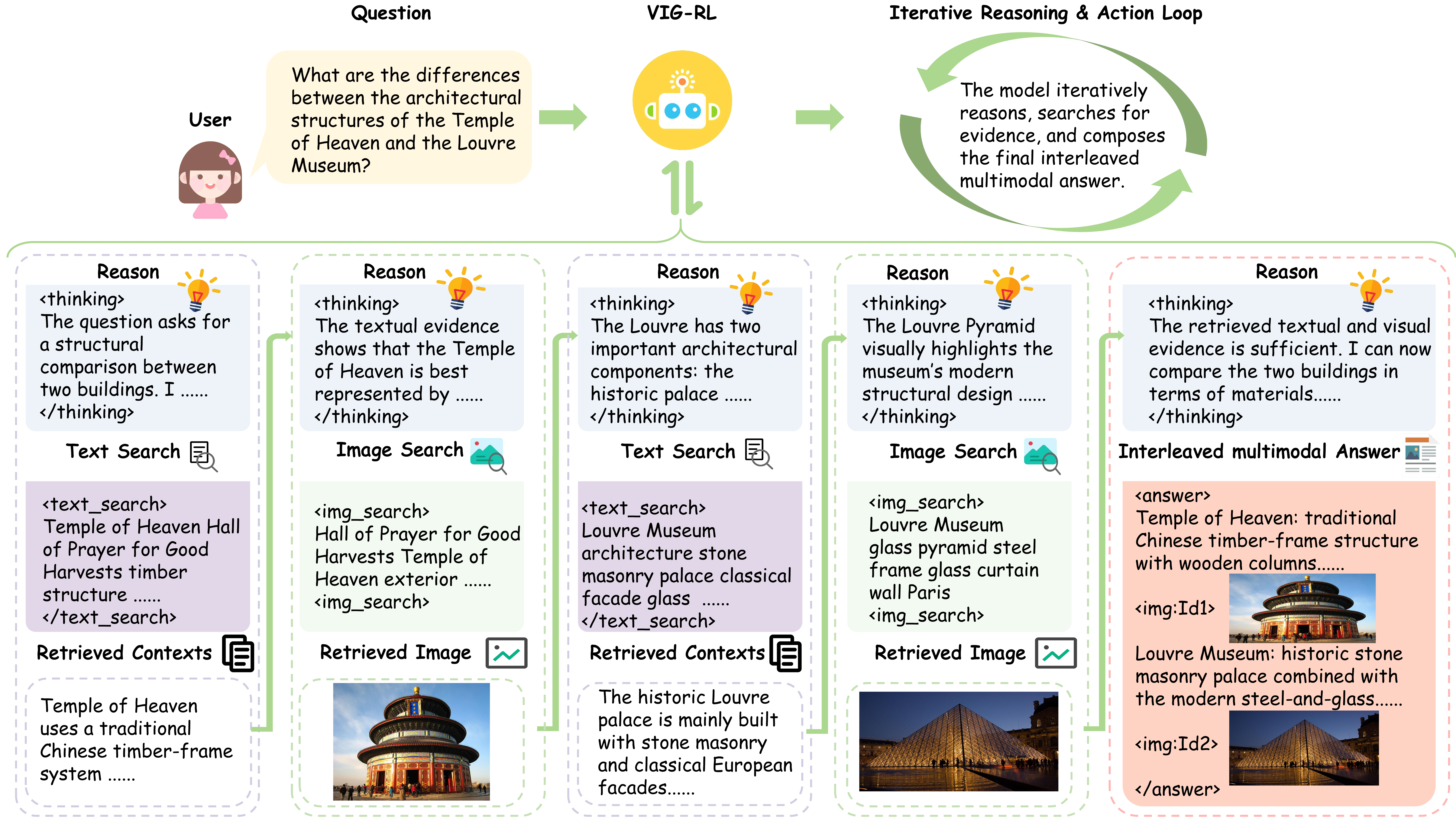}
  \caption{
Illustration of the VIG-RL workflow.
Given a user question, the agent iteratively reasons, performs text/image search, retrieves verified evidence, and composes an interleaved multimodal answer.
}
    \label{fig:main_pip}
\end{figure*}

\paragraph{Agentic Reinforcement Learning.}
Reasoning models such as OpenAI-o1~\citep{jaech2024openai} and DeepSeek-R1~\citep{guo2025deepseek} have spurred RL-based agentic reasoning. Search-R1~\citep{jin2025search} and R1-Searcher~\citep{song2025r1} train LLMs to interact with search engines, while MMSearch-R1~\citep{wu2025mmsearch} and DeepEyes V2~\citep{hong2025deepeyesv2} extend this paradigm to visual retrieval. VRAG-RL~\citep{wang2025vrag} and Vision-DeepResearch~\citep{huang2026vision} further support iterative visual exploration through operations such as cropping and region-level search, but ultimately produce text-only answers. In contrast, VIG requires agents to retrieve, select, and insert visual evidence into the final interleaved response.

\section{Method}

\subsection{Problem Formulation}

We formally define the task of Verified Image Grounding (VIG). Given a text query $q$, the objective is to generate an interleaved text-image response $A = (t_1, i_1, \dots, t_n)$, where $t_k$ is a text segment and $i_k$ is an inserted image. To ensure factual authenticity and preclude visual hallucinations, VIG strictly prohibits generative image synthesis. Instead, it imposes a hard provenance constraint: any inserted image must originate from an externally retrieved candidate pool, denoted as $\forall i_k \in A, i_k \in \mathcal{I}_{\text{retrieved}}$. The core challenge shifts from pixel-level synthesis to learning an optimal agentic policy that autonomously searches, selects, and inserts verified visual evidence.

\subsection{Agentic Interaction Framework}

To solve the VIG task, we formulate the evidence gathering and response generation as a sequential decision-making process driven by a ReAct-style \citep{yao2022react} \textit{thinking-then-acting} paradigm. Let $S_t$ denote the accumulated context state at step $t$, initialized with the user query ($S_0 = q$). At each step, the agent's policy $\pi_{\theta}$ first generates a reasoning trace (thinking) to analyze current information gaps: $r_t \sim \pi_{\theta}(\cdot \mid S_t)$. Conditioned on this reasoning, the agent samples an executable action: $a_t \sim \pi_{\theta}(\cdot \mid S_t, r_t)$ from a predefined action space $\mathcal{A}$. The environment executes $a_t$, returns an observation $o_t$, and appends it to the context for the next step: $S_{t+1} = S_t \oplus \langle r_t, a_t, o_t \rangle$, where $\oplus$ denotes sequence concatenation.

To empower the agent to navigate multimodal documents autonomously, we define the action space as $\mathcal{A} = \{a_{\text{txt}}(\cdot), a_{\text{img}}(\cdot), a_{\text{ans}}\}$:
\begin{enumerate}
    \item 
    \textbf{Text Search $a_{\text{txt}}(q_{\text{txt}})$:} The agent formulates a text search query $q_{\text{txt}}$. The environment invokes a text retriever to search an external corpus, returning an observation $o_t$ containing the top-$K_{\text{txt}}$ relevant textual passages.
    \item 
    \textbf{Image Search $a_{\text{img}}(q_{\text{img}})$:} The agent generates an image search query $q_{\text{img}}$. The environment calls a visual retriever to fetch the top-$K_{\text{img}}$ visual assets from multimodal documents. The observation $o_t$ presents each retrieved item to the MLLM as the raw image, its associated textual context, and a unique identifier (e.g., \texttt{<img:id>}). The identifier is also recorded in the candidate pool $\mathcal{I}_{\text{retrieved}}$ for subsequent selection and insertion.
    
    % The observation $o_t$ returns these images (typically represented via dense captions and identifiers), which are simultaneously recorded into the candidate pool $\mathcal{I}_{\text{retrieved}}$.
    \item 
    \textbf{Answer Generation $a_{\text{ans}}$:} Once the reasoning trace $r_t$ concludes that sufficient evidence has been aggregated, the agent triggers $a_{\text{ans}}$. This action explicitly terminates the search loop. The agent then utilizes the terminal state $S_T$ to autoregressively decode the final interleaved response $A \sim \pi_{\theta}(\cdot \mid S_T)$.
\end{enumerate}

\paragraph{Symbolic Visual Referencing.}
To bypass the modality gap, since LLMs cannot natively output visual pixels, we implement symbolic referencing to guarantee factual authenticity. The environment maps each retrieved image in $\mathcal{I}_{\text{retrieved}}$ to a unique discrete identifier (e.g., \texttt{<img:id>}). During final decoding, the policy simply predicts these symbolic tags to indicate spatial placements within the text stream. These tags are then deterministically replaced with the original source images during rendering, circumventing pixel-level hallucination. 

Additionally, to prevent infinite execution, the interaction loop strictly terminates either upon invoking $a_{\text{ans}}$ or reaching a maximum horizon $T_{\text{max}}$.

\subsection{Training the agent via GRPO and Composite Rewards}

To endow the MLLM with autonomous agentic capabilities, we optimize it via reinforcement learning using the Group Relative Policy Optimization (GRPO) algorithm~\citep{shao2024deepseekmath}. To holistically supervise the entire trajectory (from intermediate exploration $S_t$ to the terminal generation $A$) we design a composite reward system comprising four sub-components as follows:

\paragraph{Format Reward $r_{\text{format}}$.}

This reward ensures the rollout trajectory strictly conforms to the predefined interaction syntax, which acts as a binary indicator: $r_{\text{format}} = 1$ if the rollout trajectory correctly alternates between reasoning traces $r_t$ and actions $a_t$ according to our ReAct-style framework, and all tool invocations are structurally parsable; otherwise, $r_{\text{format}} = 0$.

\paragraph{Textual Outcome Reward $r_{\text{txt,LLM}}$.}

Unlike traditional QA or VQA tasks in~\citet{jin2025search,wu2025mmsearch} where Exact Match (EM) serves as a reliable metric, the open-ended nature of VIG responses renders rule-based text matching excessively sparse. Since this specific reward focuses exclusively on the natural language quality, we isolate the purely textual content $A_{\text{txt}}$ from the generated interleaved response $A$. We then employ a model-based evaluator $\pi_{\text{judge}}$ to assess the semantic alignment and factual correctness of $A_{\text{txt}}$ against the textual reference $A^*_{\text{txt}}$:$ r_{\text{txt,LLM}} \sim \pi_{\text{judge}}(\cdot \mid q, A^*_{\text{txt}}, A_{\text{txt}}). $
The judge's raw score is linearly normalized to the interval $[0,1]$ before being used as the reward signal.

\paragraph{Image Insertion Reward $r_{\text{img,pre}}$.}

To supervise the inserted image quality, employing a model-based judge (akin to $r_{\text{text}}$) makes the policy highly susceptible to reward hacking. Inspired by DeepSeek-R1~\citep{guo2025deepseek}, we combat this using a verifiable, rule-based outcome reward. Let $\mathcal{I}_{\text{ans}}$ denote the explicit image identifiers inserted in $A$. We compute a strict precision score against the ground-truth set $\mathcal{I}^*$.
Specifically, we set $r_{\text{img,pre}}=1$ when
$\mathcal{I}_{\text{ans}}=\emptyset$ and $\mathcal{I}^{*}=\emptyset$,
$r_{\text{img,pre}}=0$ when
$\mathcal{I}_{\text{ans}}=\emptyset$ and $\mathcal{I}^{*}\neq\emptyset$,
and otherwise compute
$r_{\text{img,pre}}
=|\mathcal{I}^{*}\cap\mathcal{I}_{\text{ans}}|
/|\mathcal{I}_{\text{ans}}|$.
This objective metric explicitly penalizes the indiscriminate insertion of irrelevant visual assets, anchoring the agent's multimodal alignment without relying on exploitable evaluators.

\paragraph{Search Process Reward $r_{\text{search}}$.}

To provide dense supervision over the intermediate retrieval process, we compute the recall of the ground-truth image set $\mathcal{I}^*$ within the accumulated candidate pool $\mathcal{I}_{\text{retrieved}}$: $ r_{\text{search}} = {|\mathcal{I}^* \cap \mathcal{I}_{\text{retrieved}}|}/{|\mathcal{I}^*|}. $
This explicitly encourages the policy to actively explore and gather highly relevant visual evidence.

\paragraph{Gated Composite Reward.}

Finally, these signals are aggregated into a unified reward $\mathcal{R}$:$$ \mathcal{R} = r_{\text{format}} \cdot (r_{\text{txt,LLM}} + r_{\text{img,pre}} + r_{\text{search}}). $$ The multiplicative $r_{\text{format}}$ acts as a strict structural gate, completely zeroing out rewards for invalid trajectories. This forces the agent to master the ReAct execution syntax before optimizing for reasoning and multimodal content quality.

\section{Experiments}

\subsection{Experimental Settings}

% \paragraph{Baselines.} We compare VIG-RL against state-of-the-art MLLMs under two baselines for interleaved multimodal answering.
% \textbf{(1) RAG Workflow:} This setting adopts a fixed retrieve-then-generate workflow. Specifically, it first retrieves external textual and visual evidence, and subsequently feeds the retrieved context alongside the user query to the model. Under this setting, we evaluate GPT-5~\citep{GPT-5}, Gemini-2.5-Flash~\citep{comanici2025gemini}, and five specific Qwen variants (Qwen3-VL-30B-A3B-Instruct/Thinking, Qwen3-VL-8B-Instruct/Thinking and Qwen3-VL-4B-Instruct)~\citep{bai2025qwen3}, which are instructed to directly generate the final interleaved text-image response. Furthermore, we include M2IO-R1-3B~\citep{xiao2025m2io}, which relies on a task decomposition strategy: it first generates a purely textual response and then employs a specialized, independently trained image inserter to ground the visual assets.
% \textbf{(2) Agentic Search:} To isolate the benefits of our RL training, we integrate the aforementioned closed- and open-source models into our iterative framework, functioning as active search agents that rely exclusively on zero-shot prompting.
\paragraph{Baselines.}
We compare VIG-RL with strong MLLMs and prior VIG systems under two settings.
\textbf{(1) Static RAG Workflow:} A fixed retrieve-then-generate pipeline retrieves textual and visual evidence once and provides it with the query for direct interleaved response generation. We evaluate GPT-5~\citep{GPT-5}, Gemini-2.5-Flash~\citep{comanici2025gemini}, and five Qwen3-VL variants---30B-A3B-Instruct/Thinking, 8B-Instruct/Thinking, and 4B-Instruct~\citep{bai2025qwen3}. We also include M2IO-R1-3B~\citep{xiao2025m2io}, which generates a text-only answer before inserting images with a separately trained module.
\textbf{(2) Agentic Search:} We deploy the same general-purpose MLLMs as zero-shot agents in the VIG-RL workflow, with identical retrieval tools and action space but no additional training.
% \textbf{VIG-RL (Ours):} Unlike prompt-driven baselines, our model is natively aligned via GRPO to autonomously master the complete reasoning and search-selection-insertion workflow.

\paragraph{SFT Baseline and Retrieval-Depth Selection.}
To isolate the benefit of RL from exposure to agentic demonstrations, we train an SFT baseline from the same Qwen3-VL-8B-Instruct initialization on the same 1.1k instances. For each instance, we sample five Gemini-2.5-Flash rollouts under the same agentic workflow and retain the one with the highest final-answer Image F1, yielding 1.1k trajectories for fine-tuning with LLaMA-Factory~\citep{zheng2024llamafactory}. Thus, SFT and RL share the base model, training instances, action space, and retrieval environment, differing only in offline imitation versus reward-driven optimization. For static RAG, we tune $K\in\{2,5,10\}$ and use $K=5$ by default, as it performs best overall. Each agentic search call also returns the top-5 images, while the agent adaptively decides whether and when to retrieve again.

\paragraph{Implementation Details.}
We use BGE-M3~\citep{bge-m3} for both text and image retrieval. Instead of direct cross-modal matching (e.g., CLIP), our \textit{Context-Anchored Image Retrieval} associates each image $i_j$ with its surrounding text $t_j$, forming $\mathcal{C}=\{(t_j,i_j)\}$. Given an image query $q_{\text{img}}$, we retrieve the images whose anchor texts have the highest cosine similarity to the query:
\begin{small}
\[
\mathcal{I}_{\text{retrieved}}
=
\left\{
i_j \,\middle|\,
(t_j,i_j)\in
\operatorname{TopK}_{(t,i)\in\mathcal{C}}
\cos\!\big(\mathcal{E}(q_{\text{img}}),\mathcal{E}(t)\big)
\right\},
\]
\end{small}
where $\mathcal{E}$ is the BGE-M3 text encoder. This avoids the semantic mismatch of direct visual embeddings, as examined in our ablation.
% Using Qwen3-VL-4B-Instruct and Qwen3-VL-8B-Instruct as the base models, we train VIG-RL-4B and VIG-RL-8B, respectively. We implement reinforcement learning with veRL~\citep{sheng2024hybridflow,wu2025mmsearch} and optimize both models using GRPO for $20$ epochs ($720$ optimization steps).
% For textual outcome reward, we apply Qwen3-8-Instruct as our judge model.
% The training set comprises $1.1$k curated samples, derived from an $8:2$ train-test split of the \texttt{Web}, \texttt{Wiki}, and \texttt{Arxiv} subsets of MRAMG-Bench~\citep{yu2025mramg}, inspected for cross-modal consistency. 
Using Qwen3-VL-4B/8B-Instruct as base models, we train VIG-RL-4B/8B with GRPO in veRL~\citep{sheng2024hybridflow,wu2025mmsearch} for $20$ epochs ($720$ steps). Qwen3-8B-Instruct serves as the textual-reward judge. Training uses $1.1$k cross-modally consistent samples from an $80/20$ split of the \texttt{Web}, \texttt{Wiki}, and \texttt{Arxiv} subsets of MRAMG-Bench~\citep{yu2025mramg}. To prevent retrieval-evidence leakage, the training and test sets contain no overlap in the textual chunks or images required to answer their questions.

\begin{table*}[t]
\renewcommand{\arraystretch}{1.0}
  \centering
  \resizebox{\textwidth}{!}{
    \begin{tabular}{l|cc|cc|cc|cc|ccc|ccc}
     \toprule
    \multicolumn{1}{c|}{\multirow{3}[3]{*}{\textbf{Model}}} & \multicolumn{6}{c|}{\textbf{In-Domain}}       & \multicolumn{8}{c}{\textbf{Out-of-Domain}} \\
\cmidrule{2-15}          & \multicolumn{2}{c|}{\textbf{Web}} & \multicolumn{2}{c|}{\textbf{Wiki}} & \multicolumn{2}{c|}{\textbf{Arxiv}} & \multicolumn{2}{c|}{\textbf{Wit}} & \multicolumn{3}{c|}{\textbf{Manual}} & \multicolumn{3}{c}{\textbf{Recipe}} \\
          & \textbf{F1} & \textbf{C.S.} & \textbf{F1} & \textbf{C.S.} & \textbf{F1} & \textbf{C.S.} & \textbf{F1} & \textbf{C.S.} & \textbf{F1} & \textbf{C.S.} & \textbf{Order} & \textbf{F1} & \textbf{C.S.} & \textbf{Order} \\
    \midrule
    \rowcolor{groupgray}
\multicolumn{15}{c}{\emph{\textbf{RAG Workflow}}} \\
    \midrule
    GPT-5 & \underline{96.0}  & 85.7  & 83.2  & 74.0  & 75.6  & \underline{71.5}  & 84.3  & 87.7  & 40.6  & \underline{42.7}  & \underline{30.0}  & 54.9  & \underline{55.5}  & \textbf{45.0} \\
    Gemini-2.5-Flash & 95.9  & 86.4  & 94.1  & 80.9  & \underline{83.1}  & 67.3  & 94.2  & 93.1  & 32.9  & 41.6  & 24.3  & 40.5  & 44.1  & 30.4  \\
    Qwen3-VL-30B & 90.2  & 78.3  & 90.8  & \underline{83.4}  & 69.4  & 53.4  & \textbf{97.0} & 91.4  & 31.5  & 32.2  & 21.5  & 25.5  & 30.8  & 15.2  \\
    Qwen3-VL-30B-Thinking & 92.0  & \underline{86.8}  & 81.2  & 81.7  & 70.3  & 64.6  & \textbf{97.0} & 94.3  & 31.6  & 37.6  & 22.3  & 45.9  & 54.4  & 39.1  \\
    Qwen3-VL-8B & 76.5  & 74.6  & 81.5  & 70.9  & 65.7  & 60.0  & 92.5  & 87.2  & 14.4  & 21.0  & 7.1   & 15.3  & 24.8  & 8.4  \\
    Qwen3-VL-8B-Thinking & 83.2  & 81.8  & 63.4  & 65.2  & 68.1  & 55.9  & 95.6  & 93.4  & 22.1  & 31.5  & 16.7  & 22.4  & 33.6  & 17.4  \\
    Qwen3-VL-4B & 48.3  & 53.4  & 63.4  & 60.0  & 43.2  & 45.9  & 95.4  & 88.3  & 10.4  & 21.0  & 3.5   & 5.4   & 22.7  & 1.0  \\
    M2IO-R1-3B & -     & -     & -     & -     & 72.0  & -     & -     & -     & 40.0  & -     & 29.1  & 51.2  & -     & 31.1  \\
    \midrule
        \rowcolor{groupgray}
\multicolumn{15}{c}{\emph{\textbf{Agentic Search}}} \\
    \midrule
    GPT-5 & 71.2  & 55.3 & 30.7  & 56.9 & 20.3  & 35.4 & 18.2  & 40.9 & \textbf{44.7}  & 32.5  & 28.8  & 40.5  & 41.4  & 26.9  \\
    Gemini-2.5-Flash & 79.5  & 66.2  & 46.5  & 58.0  & 23.3  & 34.9  & 39.9  & 68.1  & 34.3  & 34.2  & 24.1  & 43.4  & 47.2  & 28.4  \\
    Qwen3-VL-30B & 58.9  & 53.0  & 18.4  & 35.8  & 38.4  & 39.5  & 44.5  & 53.7  & 19.9  & 22.5  & 12.1  & 16.6  & 24.0  & 8.0  \\
    Qwen3-VL-30B-Thinking & 60.8  & 55.9  & 42.1  & 47.9  & 51.3  & 41.5  & 69.8  & 68.7  & 28.0  & 28.5  & 12.8  & 24.7  & 32.4  & 16.0  \\
    Qwen3-VL-8B-Thinking & 36.2  & 37.3  & 35.6  & 42.9  & 20.8  & 25.9  & 31.4  & 44.5  & 12.7  & 21.5  & 5.8   & 12.6  & 23.2  & 6.3  \\
    \midrule
        \rowcolor{groupgray}
\multicolumn{15}{c}{\emph{\textbf{Ours}}} \\
    \midrule
    Qwen3-VL-4B (Agentic) & 47.8  & 42.5  & 20.5  & 37.2  & 18.8  & 31.0  & 20.4  & 41.4  & 14.7  & 20.0  & 7.8   & 14.5  & 24.4  & 6.6  \\
     \rowcolor{blue1}
    VIG-RL-4B (Ours)  & 95.6  & 80.7  & \underline{94.7}  & 75.5  & 80.6  & 63.9  & 95.3  & 86.5  & 40.5  & \underline{42.7}  & 29.0  & \textbf{55.3} & 46.3  & 35.1  \\
    \color{softgreen}$\Delta$      & \color{softgreen}\textbf{+47.8}  & \color{softgreen}\textbf{+38.2}  & \color{softgreen}\textbf{+74.3}  & \color{softgreen}\textbf{+38.3}  & \color{softgreen}\textbf{+61.8}  & \color{softgreen}\textbf{+32.9}  & \color{softgreen}\textbf{+74.9} & \color{softgreen}\textbf{+45.1}  & \color{softgreen}\textbf{+25.8}  & \color{softgreen}\textbf{+22.7}  & \color{softgreen}\textbf{+21.2}  & \color{softgreen}\textbf{+40.8}  & \color{softgreen}\textbf{+21.9}  & \color{softgreen}\textbf{+28.5}  \\
    \midrule
    Qwen3-VL-8B (Agentic) & 50.4  & 46.7  & 36.6  & 46.5  & 24.1  & 36.1  & 29.5  & 44.3  & 22.9  & 24.1  & 12.4  & 20.0  & 27.3  & 10.9  \\
     \rowcolor{blue1}
    VIG-RL-8B (Ours)  & \textbf{97.2} & \textbf{94.0} & \textbf{95.7} & \textbf{88.9} & \textbf{84.2} & \textbf{81.7} & \underline{96.7}  & \textbf{94.5} & \underline{44.3} & \textbf{48.1} & \textbf{32.2} & \underline{55.1}  & \textbf{61.4} &  \underline{40.9} \\
    \color{softgreen}$\Delta$ & \color{softgreen}\textbf{+46.8} & \color{softgreen}\textbf{+47.3}  & \color{softgreen}\textbf{+59.1}  & \color{softgreen}\textbf{+42.4} & \color{softgreen}\textbf{+60.1} & \color{softgreen}\textbf{+45.6}  & \color{softgreen}\textbf{+67.2}  & \color{softgreen}\textbf{+50.2}  & \color{softgreen}\textbf{+21.4}  & \color{softgreen}\textbf{+24.0} & \color{softgreen}\textbf{+19.8}  & \color{softgreen}\textbf{+35.0} & \color{softgreen}\textbf{+34.1}  & \color{softgreen}\textbf{+30.0}  \\
    \bottomrule
    \end{tabular}%
}
  %   \caption{Main results on MRAMG-Bench. Performance comparison of VIG-RL and state-of-the-art LMMs across six datasets under RAG Workflow and Agentic Search paradigms. F1 denotes the image-level F1 score of the inserted images in the final answer. C.S. denotes the Comprehensive Score, which measures the overall quality of the interleaved multimodal response. The best results are shown in \textbf{bold}, and the second best are \underline{underlined}.}
  % \label{main_res}%
  \caption{
% Main results across six datasets from MRAMG-Bench under static RAG Workflow and Agentic Search settings. F1 evaluates image-set selection accuracy, while C.S. evaluates the overall interleaved response, including textual correctness, image relevance, semantic placement, text--image alignment, and multimodal coherence. Order explicitly evaluates the relative sequence of inserted images on the order-sensitive \texttt{Manual} and \texttt{Recipe} datasets. VIG-RL-4B provides a similar-scale comparison with M2IO-R1-3B. \textcolor{softgreen}{$\Delta$} denotes improvement over the corresponding agentic backbone, computed from unrounded scores.
% The best results are shown in \textbf{bold}, and the second best are \underline{underlined}.
Main results on six MRAMG-Bench datasets under static RAG and agentic search. Image F1 measures image selection; C.S. jointly evaluates text quality, image relevance, semantic placement, text--image alignment, and multimodal coherence; and Order measures image sequence on \texttt{Manual} and \texttt{Recipe}. \textcolor{softgreen}{$\Delta$} denotes gains over the corresponding zero-shot agentic baseline, computed from unrounded scores. Best and second-best results are \textbf{bolded} and \underline{underlined}.
}
\label{tab:main-results}
\end{table*}%

\paragraph{Benchmarks.}
We evaluate on all six datasets of MRAMG-Bench~\citep{yu2025mramg}, one of the few publicly available retrieval-grounded VIG benchmarks with explicit ground-truth interleaved answers. Spanning Web, academic, and lifestyle domains, they provide broad coverage beyond a single-domain evaluation. We evaluate \textbf{In-Domain} performance on the held-out splits of \texttt{Web}, \texttt{Wiki}, and \texttt{Arxiv}, and \textbf{OOD} generalization on \texttt{Wit}, \texttt{Recipe}, and \texttt{Manual}.

\paragraph{Metrics.}
Our metrics jointly evaluate \emph{which} images are inserted, \emph{where}, and \emph{in what order}. \textbf{Image F1} measures set-level selection precision and recall; unlike the precision-based training reward, it also captures visual-evidence coverage but ignores placement. \textbf{Comprehensive Score (C.S.)} uses GPT-4o~\citep{gpt4o} to assess textual correctness, image relevance, insertion-position appropriateness, text--image alignment, and overall coherence, thereby explicitly evaluating \emph{where to insert}. On the procedural \texttt{Manual} and \texttt{Recipe} datasets, \textbf{Order Score}~\citep{yu2025mramg} additionally measures the relative image order via weighted edit distance.

\subsection{Main Results}
\label{sec:main-results}

% \paragraph{Overall Performance.}
% Table~\ref{tab:main-results} reports the results across six MRAMG-Bench datasets. 
% Our VIG-RL-8B performs consistently well across datasets and metrics, achieving the best C.S. on all six datasets together with strong Image F1 results across domains.
% These gains show that tool access and prompting alone do not produce a reliable VIG agent; reinforcement learning is essential for acquiring an effective search-selection-insertion policy.

\paragraph{Overall Performance.}
Table~\ref{tab:main-results} reports results across the six MRAMG-Bench datasets.
VIG-RL-8B achieves the strongest overall performance, attaining the best C.S. on all six datasets together with consistently strong Image F1 and Order scores.
Its robust performance across image selection, semantic placement, and sequence ordering demonstrates that VIG-RL successfully learns an effective search--selection--insertion policy.

\paragraph{VIG-RL Surpasses Static RAG.}
VIG-RL-8B consistently outperforms strong static RAG systems in overall response quality.
Relative to the same Qwen3-VL-8B backbone under static RAG, it improves the six-dataset macro-average Image F1 from $57.7$ to $78.9$ (\textbf{$+21.2$}) and C.S. from $56.4$ to $78.1$ (\textbf{$+21.7$}), showing that the gains cannot be attributed merely to the underlying model checkpoint.
Even against a conservative per-dataset oracle that selects the strongest static RAG system for each benchmark, VIG-RL-8B improves macro-averaged C.S. from $72.4$ to $78.1$ (\textbf{$+5.7$}).
The advantage remains pronounced when image selection is already accurate: on \texttt{Web}, RAG-based GPT-5 obtains $96.0$ Image F1 but only $85.7$ C.S., whereas VIG-RL-8B reaches $97.2$ F1 and $94.0$ C.S.
This large C.S. gain despite similar Image F1 indicates improvements beyond set-level selection, particularly in semantic placement, text--image alignment, and overall multimodal coherence.

\paragraph{Zero-Shot Generalization and Ordered Image Grounding.}
On the unseen \texttt{Wit}, \texttt{Manual}, and \texttt{Recipe} domains, VIG-RL-8B achieves the best C.S. of $94.5$, $48.1$, and $61.4$, respectively, demonstrating strong zero-shot transfer. This is particularly notable on the order-sensitive \texttt{Manual} and \texttt{Recipe} datasets, where successful grounding requires both selecting relevant images and organizing them according to the underlying procedural structure. VIG-RL-8B achieves the best Order Score on \texttt{Manual} ($32.2$ vs.\ $30.0$ for the strongest baseline) and the second-best Image F1 and Order Score on \texttt{Recipe} ($55.1$ and $40.9$), while attaining the best C.S. These results show that the learned policy transfers across domains while preserving both semantic grounding and the relative organization of visual evidence.

% \paragraph{Effectiveness at the 4B Scale.}
% The gains of VIG-RL persist at a smaller model scale. Compared with its zero-shot Qwen3-VL-4B agentic backbone, VIG-RL-4B improves every reported metric, including Image F1 from 18.8 to 80.6 and C.S. from 31.0 to 63.9 on \texttt{Arxiv}. It also outperforms the similarly sized M2IO-R1-3B on four of five reported metrics, improving \texttt{Arxiv} F1 from 72.0 to 80.6 and \texttt{Recipe} F1/Order from 51.2/31.1 to 55.3/35.1. These results show that the learned policy remains effective without relying on the larger 8B backbone.

\paragraph{Effectiveness at the 4B Scale.}
VIG-RL remains effective at a smaller model scale. Compared with the corresponding zero-shot Qwen3-VL-4B agentic baseline, VIG-RL-4B improves every metric, including Image F1 from $18.8$ to $80.6$ and C.S. from $31.0$ to $63.9$ on \texttt{Arxiv}. It also outperforms the similarly sized M2IO-R1-3B on four of the five metrics available for comparison, improving \texttt{Arxiv} F1 from $72.0$ to $80.6$ and \texttt{Recipe} F1/Order from $51.2/31.1$ to $55.3/35.1$. These results demonstrate that the effectiveness of VIG-RL is not confined to the larger 8B model.

% {\color{red} \#binghui: M2IO why not all six datasets?}

% \begin{figure}[t]
% \centering
% \includegraphics[width=\linewidth]{latex/Fig/sft.pdf}

% \caption{
% % Comparison of zero-shot agentic prompting (Base), supervised trajectory imitation (SFT), and reinforcement learning (VIG-RL). Bars report the Comprehensive Score, while the dashed lines denote the absolute improvement over the corresponding base model. SFT and VIG-RL use the same base model and the same 1.1k training instances.
% Comparison of the zero-shot agentic baseline (Base), supervised trajectory imitation (SFT), and VIG-RL. Bars show C.S., and dashed lines show absolute gains over Base. SFT and VIG-RL share the same initialization and 1.1k training instances.
% }
% \label{fig:sft-vs-rl}
% \end{figure}

\subsection{Further Analysis}
\label{sec:further-analysis}

% \paragraph{RL versus Supervised Trajectory Imitation.}

% To determine whether the gains of VIG-RL can be reproduced through supervised imitation, we compare three variants initialized from the same Qwen3-VL-8B-Instruct model: the zero-shot agentic baseline (Base), the trajectory-supervised model (SFT), and VIG-RL. The SFT model is trained on search-and-answer trajectories generated by Gemini-2.5-Flash using the same 1.1k training instances and agentic workflow as VIG-RL, providing a strong training-instance-matched imitation baseline.
% As shown in Figure~\ref{fig:sft-vs-rl}, SFT improves over the zero-shot agentic model, confirming that  expert trajectories provide useful supervision. 
% Nevertheless, VIG-RL performs better on all six datasets, with clear gains on \texttt{Arxiv} (81.7 vs.\ 41.8 C.S.), \texttt{Manual} (48.1 vs.\ 33.0), and \texttt{Recipe} (61.4 vs.\ 48.4). These results show that offline trajectory imitation alone is insufficient for reliable VIG: unlike SFT, VIG-RL learns from its own interactions and receives direct feedback on both evidence acquisition and final grounding quality, enabling it to adjust its decisions as the search process evolves.

\begin{figure}[t]
\centering
\includegraphics[width=\linewidth]{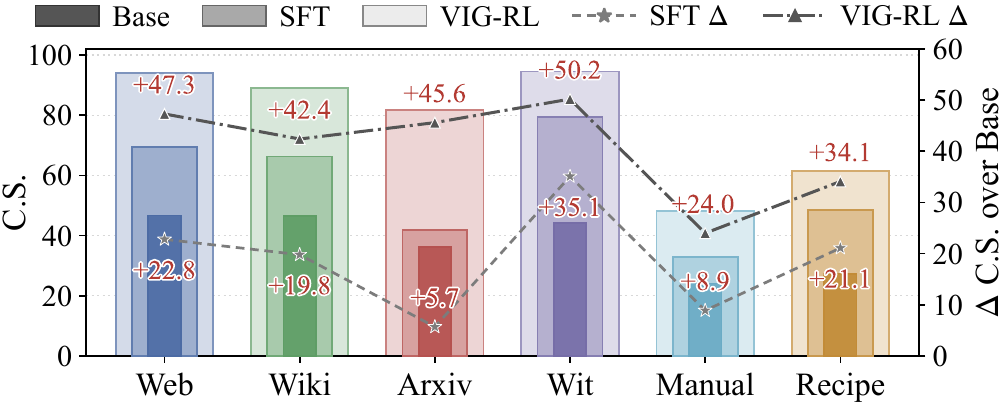}

\caption{
% Comparison of zero-shot agentic prompting (Base), supervised trajectory imitation (SFT), and reinforcement learning (VIG-RL). Bars report the Comprehensive Score, while the dashed lines denote the absolute improvement over the corresponding base model. SFT and VIG-RL use the same base model and the same 1.1k training instances.
Comparison of the zero-shot agentic baseline (Base), supervised trajectory imitation (SFT), and VIG-RL. Bars show C.S., and dashed lines show absolute gains over Base. SFT and VIG-RL share the same initialization and 1.1k training instances.
}
\label{fig:sft-vs-rl}
\end{figure}

\paragraph{RL versus Supervised Trajectory Imitation.}
We compare three variants initialized from Qwen3-VL-8B-Instruct: the zero-shot agentic baseline (Base), SFT, and VIG-RL. SFT uses best-of-five Gemini-2.5-Flash trajectories generated from the same 1.1k instances and agentic workflow as VIG-RL, providing a strong matched imitation baseline. Figure~\ref{fig:sft-vs-rl} shows that SFT consistently improves over Base, confirming the value of trajectory supervision. Nevertheless, VIG-RL achieves higher C.S. on all six datasets, with large gains over SFT on \texttt{Arxiv} ($81.7$ vs.\ $41.8$), \texttt{Manual} ($48.1$ vs.\ $33.0$), and \texttt{Recipe} ($61.4$ vs.\ $48.4$). Unlike SFT's imitation of fixed demonstrations, VIG-RL optimizes its own interactions using rewards for evidence acquisition and final grounding, jointly adapting its search, selection, and insertion decisions.

% \paragraph{Adaptive Retrieval beyond Fixed-Depth RAG.}

% Static RAG must balance evidence coverage against the context noise introduced by irrelevant candidates. To ensure that the static baseline is not disadvantaged by an arbitrarily small candidate pool, we evaluate \(K\in\{2,5,10\}\) using Qwen3-VL-8B-Instruct as a representative RAG backbone. As shown in Figure~\ref{fig:topk-ablation}, performance first improves and then declines as \(K\) increases. On \texttt{Arxiv}, F1/C.S. rises from 55.5/51.7 at \(K=2\) to 65.7/60.0 at \(K=5\), but drops to 50.9/50.2 at \(K=10\). The same trend appears on \texttt{Recipe}, where \(K=5\) performs best and \(K=10\) degrades all three metrics. We therefore use \(K=5\) in the main RAG comparisons.
% Both settings retrieve the top-$5$ candidates per call, while VIG-RL adaptively decides whether to search again. Thus, the gain is not due to a larger per-call retrieval depth, but is consistent with the benefit of adaptive retrieval.

\begin{figure}[t]
\centering
\includegraphics[width=0.9\linewidth]{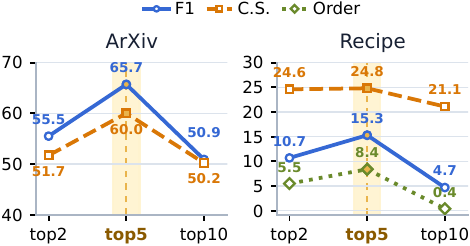}

\caption{
% Effect of the image-retrieval depth $K$ in the static RAG workflow using Qwen3-VL-8B-Instruct. A moderate retrieval depth of $K=5$ achieves the best overall performance on both \texttt{Arxiv} and \texttt{Recipe}, balancing evidence coverage against the context noise introduced by excessive candidates.
Effect of static-RAG retrieval depth $K$ with Qwen3-VL-8B-Instruct. Among the tested settings, $K=5$ performs best overall on \texttt{Arxiv} and \texttt{Recipe}, balancing evidence coverage and context noise.
}
\label{fig:topk-ablation}
\end{figure}

\paragraph{Adaptive Retrieval beyond Fixed-Depth RAG.}
For one-shot static RAG, \(K\) reflects a coverage--noise trade-off rather than a monotonically increasing information budget: too few candidates omit necessary evidence, whereas too many introduce irrelevant context. We therefore compare VIG-RL against static RAG at its empirically strongest retrieval depth, rather than matching raw candidate counts across the two workflows. Using Qwen3-VL-8B-Instruct, we tune \(K\in\{2,5,10\}\). As shown in Figure~\ref{fig:topk-ablation}, \texttt{Arxiv} F1/C.S. increases from \(55.5/51.7\) at \(K=2\) to \(65.7/60.0\) at \(K=5\), but decreases to \(50.9/50.2\) at \(K=10\); \texttt{Recipe} exhibits the same trend. We thus use \(K=5\), the strongest tested static-RAG setting, in all main comparisons. VIG-RL also retrieves top-5 candidates per call, but adaptively decides whether, when, and how to search again. Its advantage therefore reflects adaptive retrieval over a tuned fixed-depth workflow, rather than comparison with an arbitrarily small static candidate pool.

% \paragraph{VIG-RL Learns to Search, Select, and Insert.}

% To explain why prompt-driven agents still underperform static RAG, Figure~\ref{fig:agentic_retrieval_img_search} analyzes their retrieval behavior on \texttt{Wit}. Qwen3-VL-8B-Instruct and Gemini-2.5-Flash invoke image search on only $36.4\%$ and $30.6\%$ of examples, despite achieving conditional retrieval recalls of $95.5\%$ and $100.0\%$. Thus, their dominant bottleneck is failing to trigger image search
% when visual grounding is required. VIG-RL raises the search rate to $100\%$ and improves conditional Answer F1/C.S. to $96.7/94.5$, compared with $81.1/78.0$ for Qwen and $91.0/90.5$ for Gemini. These results show that VIG-RL improves both search triggering and the grounding of retrieved evidence.

\begin{figure}[t]
\centering
\includegraphics[width=\linewidth]{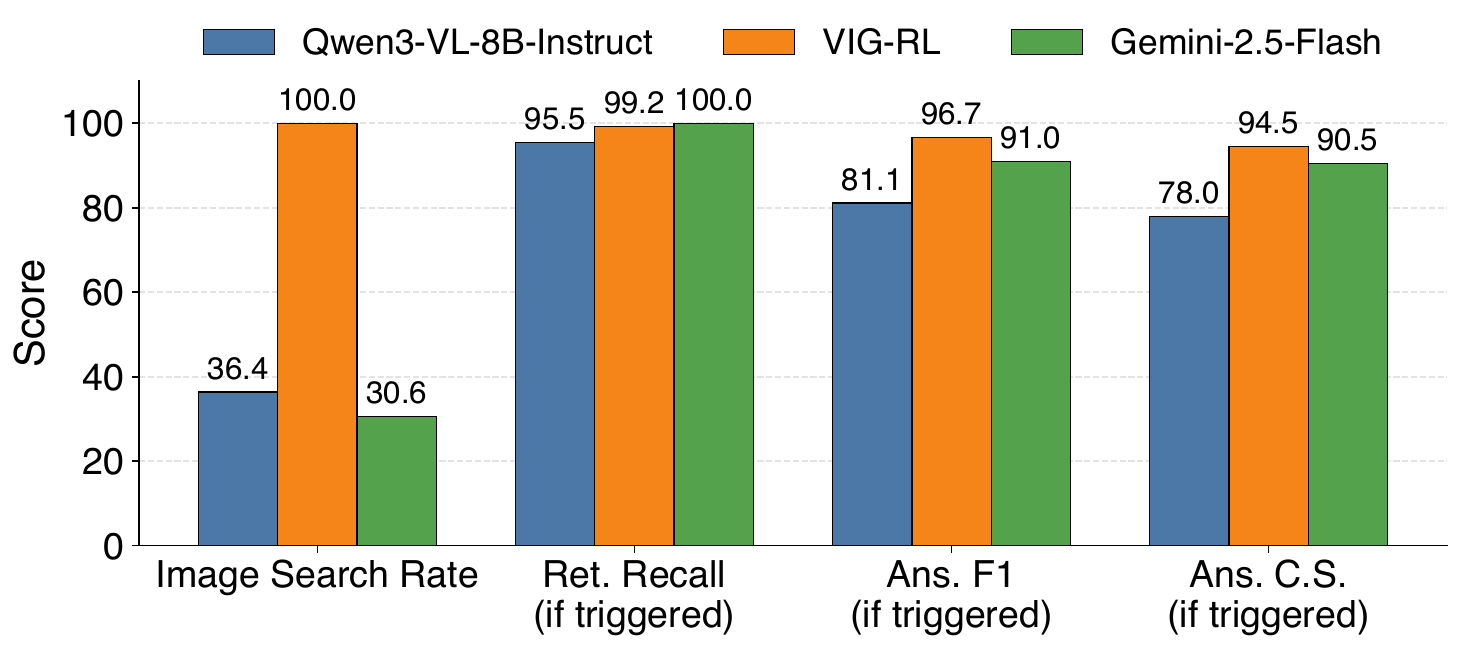}
\caption{
% Comparison of agentic retrieval behavior and downstream response quality on the \texttt{Wit} dataset.
% Image Search Rate is the proportion of examples where retrieval is triggered.
% Ret. Recall is the recall of gold verified images among retrieved images.
% Ans. F1 is the image-level F1 between inserted and gold images.
% Ans. C.S. is the Comprehensive Score of the final interleaved response.
% Except for Search Rate, all metrics are computed only on examples where retrieval is triggered.
Agentic retrieval and response quality on \texttt{Wit}. Search Rate is the fraction of examples triggering image retrieval; Ret. Recall measures the fraction of gold images retrieved; Ans. F1 and C.S. evaluate image selection and overall interleaved-response quality, respectively. All metrics except Search Rate are computed only on examples that trigger retrieval.
}
\label{fig:agentic_retrieval_img_search}
\end{figure}

\paragraph{VIG-RL Learns to Search, Select, and Insert.}
Figure~\ref{fig:agentic_retrieval_img_search} decomposes performance on \texttt{Wit} into search triggering and post-retrieval grounding. Qwen3-VL-8B-Instruct and Gemini-2.5-Flash invoke image search on only $36.4\%$ and $30.6\%$ of examples, despite conditional retrieval recalls of $95.5\%$ and $100.0\%$, indicating their main bottleneck is failing to search when visual evidence is required. VIG-RL raises the search rate to $100\%$ and improves conditional Ans. F1/C.S. to $96.7/94.5$, compared with $81.1/78.0$ for Qwen and $91.0/90.5$ for Gemini. Thus, VIG-RL learns when to retrieve and how to select and integrate the retrieved evidence, forming an effective search--selection--insertion policy.

% \begin{figure}[t]
% \centering
% \includegraphics[width=\linewidth]{latex/Fig/sft.pdf}

% \caption{
% % Comparison of zero-shot agentic prompting (Base), supervised trajectory imitation (SFT), and reinforcement learning (VIG-RL). Bars report the Comprehensive Score, while the dashed lines denote the absolute improvement over the corresponding base model. SFT and VIG-RL use the same base model and the same 1.1k training instances.
% Comparison of the zero-shot agentic baseline (Base), supervised trajectory imitation (SFT), and VIG-RL. Bars show C.S., and dashed lines show absolute gains over Base. SFT and VIG-RL share the same initialization and 1.1k training instances.
% }
% \label{fig:sft-vs-rl}
% \end{figure}

\begin{figure}[t]
\centering
\includegraphics[width=0.9\linewidth]{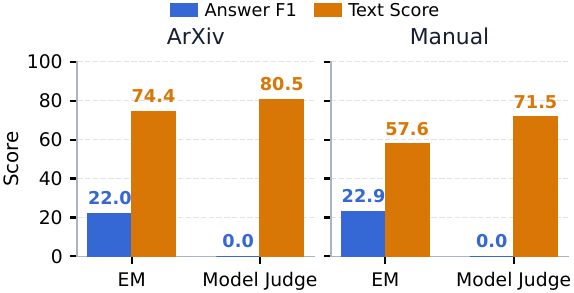}

\caption{
% Performance comparison of two reward variants (EM reward and pure model judge) on \texttt{Arxiv} and \texttt{Manual}. 
% Answer F1 measures image selection correctness, while Text Score evaluates textual quality. 
Comparison of EM and model-judge-only rewards on \texttt{Arxiv} and \texttt{Manual}. Answer F1 measures image-selection accuracy, and Text Score measures textual quality.
}
\label{fig:agentic_retrieval}
\end{figure}

\begin{table*}[htbp]
  \centering
 
  \resizebox{0.9\textwidth}{!}{
    \begin{tabular}{l|cccc|cccc}
 \toprule
 \cmidrule{2-9}    
\multicolumn{1}{c|}{\multirow{3}[3]{*}{\textbf{Variants}}} & \multicolumn{4}{c|}{\textbf{Arxiv (In-domain)}} & \multicolumn{4}{c}{\textbf{Manual (Out-of-Domain)}} \\
\cmidrule{2-9}    \multicolumn{1}{c|}{} & \multicolumn{3}{c}{\textbf{Final Answer}} & \multicolumn{1}{c|}{\textbf{Retrieval }} & \multicolumn{3}{c}{\textbf{Final Answer}} & \multicolumn{1}{c}{\textbf{Retrieval }} \\
    \multicolumn{1}{c|}{} & \multicolumn{1}{c}{\textbf{F1}} & \multicolumn{1}{c}{\textbf{C.S.}} & \multicolumn{1}{c}{\textbf{Recall}} & \multicolumn{1}{c|}{\textbf{Recall}} & \multicolumn{1}{c}{\textbf{F1}} & \multicolumn{1}{c}{\textbf{C.S.}} & \multicolumn{1}{c}{\textbf{Recall}} & \multicolumn{1}{c}{\textbf{Recall}} \\
    \midrule
    Qwen3-VL-8B-Instruct (w/o RL) & 24.1 & 36.1 & 29.3 & 29.3 & 22.9 & 24.1 & 26.8 & 36.5 \\
    \midrule
 \rowcolor{groupgray}
    \multicolumn{9}{c}{\textbf{\textit{Reward Design}}} \\
    \midrule
    $r_{\text{format}} \cdot r_{\text{EM}}$ & 22.0  & 30.5  & 24.4  & 26.8  & 22.9  & 21.0  & 25.1  & 32.1  \\
    $r_{\text{format}} \cdot r_{\text{MLLM}}$ & 0.0   & 33.4  & 0.0   & 0.0   & 0.0   & 29.1  & 0.0   & 0.0  \\
     $r_{\text{format}} \cdot (r_{\text{txt,LLM}}+r_{\text{img,pre}})$ & 77.6  & 67.6  & 76.4  & 87.8  & 34.9  & \underline{42.2}  & 32.0  & 52.9  \\
     $r_{\text{format}} \cdot (r_{\text{txt,LLM}}+r_{\text{img,rec}}+r_{\text{search}})$ & 70.7  & 50.2  & \textbf{87.8}  & 87.8  & 36.4  & 32.8  & \underline{43.1}  & 52.6  \\
     $r_{\text{format}} \cdot (r_{\text{txt,LLM}}+r_{\text{img,F1}}\,+r_{\text{search}})$ & \underline{78.0}  & 62.2  & \textbf{87.8}  & 90.2  & \underline{42.4}  & 36.6  & \textbf{44.6}  & 53.9  \\
    $r_{\text{format}} \cdot (r_{\text{MLLM}}\;+r_{\text{img,pre}}+r_{\text{search}})$ & 70.4  & \underline{68.0}  & 71.5  & \textbf{92.7}  & 29.5  & 34.3  & 25.9  & \underline{53.7}  \\
    
        \rowcolor{blue1}
        $r_{\text{format}} \cdot (r_{\text{txt,LLM}}+r_{\text{img,pre}}+r_{\text{search}})$  & \textbf{84.2}  & \textbf{81.7}  & \textbf{87.8}  & \textbf{92.7}  & \textbf{44.3}  & \textbf{48.1}  & 42.5  & \textbf{55.1}  \\
    \midrule
    \rowcolor{groupgray}
    \multicolumn{9}{c}{\textbf{\textit{Retrieval Strategy}}} \\
    \midrule
    
    {\small CLIP-Style Image Search \,\,\,\,\;\,\,\,\,\,\,\,\,\,\, +Text Search}  & 51.1  & 42.0  & 68.3  & 84.6  & 29.6  & 30.4  & 37.1  & 47.9  \\
    {\small Context-Anchored Image Search} & 63.5  & 50.5  & 73.6  & 74.4  & 38.3  & 36.5  & 39.9  & 51.8  \\
\rowcolor{blue1}
    {\small Context-Anchored Image Search +Text Search}   & \textbf{84.2}  & \textbf{81.7}  & \textbf{87.8}  & \textbf{92.7}  & \textbf{44.3}  & \textbf{48.1}  & 42.5  & \textbf{55.1}  \\
    \midrule
    \end{tabular}%
    }
     \caption{ 
     Ablation study of reward designs and retrieval strategies. Final-answer Recall measures the fraction of ground-truth images included in the generated response, while retrieval Recall measures the fraction retrieved before answer generation.}
% Ablation study of reward designs and retrieval strategies on \texttt{Arxiv} and \texttt{Manual}. %We report image-level F1, Comprehensive Score (C.S.), final-answer Recall, and retrieval Recall. 
% Final-answer Recall measures the fraction of ground-truth images correctly included in the generated response, while retrieval Recall measures the fraction of ground-truth images successfully retrieved before answer generation. }
  \label{tab:Ablation}%
\end{table*}%

\subsection{Ablation Studies.}

\subsubsection{Reward Design.}

% \paragraph{Necessity of Hybrid Reward Design.} We first demonstrate the necessity of our hybrid reward by evaluating the failure modes of standard formulations on the VIG task (Table~\ref{tab:Ablation} and Figure~\ref{fig:agentic_retrieval}). Applying a simple Exact-Match (EM) reward yields a catastrophic performance drop, achieving merely 22.0 Answer F1 and 30.5 Comprehensive Score (C.S.) on the \texttt{Arxiv} dataset. Because VIG requires long-form generation, EM rigidly penalizes valid semantic paraphrasing, providing a sparse and brittle signal. Conversely, relying on a pure LLM judge to evaluate the holistic response induces severe reward hacking. As strikingly revealed in Figure~\ref{fig:agentic_retrieval}, the pure model judge achieves a high Text Score (80.5) but a completely collapsed Answer F1 ($0.0$). The agent exploits the model's preference for textual fluency while entirely hallucinating unverified image identifiers. These dual failures strictly necessitate our hybrid paradigm, which simultaneously accommodates textual flexibility (via the model judge) and rigorously enforces visual verification (via rule-based exactness).

\paragraph{Necessity of Hybrid Reward Design.}
Table~\ref{tab:Ablation} and Figure~\ref{fig:agentic_retrieval} reveal the limitations of standard reward formulations for VIG. An Exact-Match (EM) reward yields only $22.0$ Answer F1 and $30.5$ C.S. on \texttt{Arxiv}, as exact matching penalizes valid paraphrases in long-form responses and provides a sparse, brittle signal. Conversely, a pure MLLM-judge reward is vulnerable to reward hacking: despite a high Text Score of $80.5$, Answer F1 collapses to $0.0$ because the policy favors fluent text while hallucinating image identifiers. These complementary failures motivate our hybrid reward, which combines model-based evaluation for semantic text quality with a verifiable rule-based reward for image grounding.

% \paragraph{Necessity of Process-Level Supervision.} Building on the hybrid architecture, we demonstrate that relying solely on terminal answer-level supervision is insufficient for complex multimodal retrieval. When we ablate the intermediate process reward, the agent lacks explicit guidance on active evidence acquisition. As shown in Table~\ref{tab:Ablation}, this omission directly impairs the agent's search capability, dropping Retrieval Recall from 92.7 to 87.8 on \texttt{Arxiv}. Consequently, the failure to fetch necessary visual assets cascades into a degraded final Answer F1 (dropping from 84.2 to 77.6). This proves that a dedicated process reward is essential to teach the agent \textit{how} to locate relevant contexts. 
% Furthermore, because this process signal effectively guarantees high candidate Recall during intermediate steps, the terminal answer reward only needs to focus on Precision to filter noise. Replacing our precision reward with Recall- or F1-based variants causes the agent to over-insert retrieved candidates, plummeting the Comprehensive Score to 50.2 and 62.2, respectively.

\paragraph{Necessity of Process-Level Supervision.}
Terminal supervision alone is insufficient for multimodal retrieval. As shown in Table~\ref{tab:Ablation}, removing the process reward reduces Retrieval Recall on \texttt{Arxiv} from $92.7$ to $87.8$ and consequently lowers Answer F1 from $84.2$ to $77.6$, demonstrating the value of explicitly supervising evidence acquisition. Moreover, once the process reward promotes high candidate recall, the terminal image reward should emphasize precision to suppress irrelevant insertions. Replacing it with Recall- or F1-based rewards encourages over-insertion and reduces C.S. to $50.2$ and $62.2$, respectively. These results reveal a complementary division of labor: process supervision promotes evidence coverage, while precision-based outcome reward filters noise.

\paragraph{Strict Reward Decoupling Enables Emergent Placement.}
Our reward is deliberately factorized: the process term promotes candidate coverage, the precision-based image reward filters irrelevant evidence, and the text-only judge evaluates semantic quality. Replacing the latter with a holistic MLLM judge introduces overlapping and exploitable visual supervision, reducing Answer F1/C.S. from $84.2/81.7$ to $70.4/68.0$ on \texttt{Arxiv} and from $44.3/48.1$ to $29.5/34.3$ on \texttt{Manual}; pure MLLM supervision also exhibits reward hacking (Figure~\ref{fig:agentic_retrieval}). Although final-answer recall is not directly rewarded, Recall- and F1-based alternatives encourage over-insertion and reduce C.S. to $50.2$ and $62.2$, supporting this coverage--filtering decomposition. More importantly, our reward is invariant to image placement and order: any two format-valid outputs with the same stripped text, image-ID set, and search trajectory receive identical rewards. Nevertheless, with the backbone and prompt fixed, VIG-RL raises Order Score from $12.4$ to $32.2$ on \texttt{Manual} and from $10.9$ to $40.9$ on \texttt{Recipe}. These gains therefore emerge without positional labels or reward shaping, suggesting that sequence-level RL amplifies the pretrained MLLM's latent text--image alignment prior once retrieval and selection become reliable.

\subsubsection{Retrieval Strategy}

\paragraph{Context-Anchored vs. Direct Visual Retrieval.}
We compare our context-anchored image retrieval with direct text-to-image matching using BGE-VL-Base~\citep{zhou2024megapairs}. As shown in Table~\ref{tab:Ablation}, direct retrieval reduces Retrieval Recall from $92.7$ to $84.6$ and Answer F1 from $84.2$ to $51.1$ on \texttt{Arxiv}. This indicates visual embeddings alone struggle with knowledge-intensive figures whose meanings depend heavily on captions and surrounding text. By ranking image-associated textual contexts and returning corresponding figures, our method better preserves such semantic information.

\paragraph{Text Search Supports Multi-Hop Retrieval.}
Removing text search on \texttt{Manual} reduces C.S. from $48.1$ to $36.5$ and Image Retrieval Recall from $55.1$ to $51.8$. The decline in visual recall suggests that text search not only supplies knowledge for answer generation, but also provides context for refining subsequent image queries. Text and image search therefore operate jointly as a multi-step evidence-acquisition policy.

\subsection{Case Study}

Detailed case studies are provided in Appendix. On the \texttt{Manual} dataset, Qwen3-VL-8B-Instruct (Figure~\ref{fig:case2}) retrieves text but skips image search, producing a text-only response, whereas Gemini-2.5-Flash (Figure~\ref{fig:case3}) selects an irrelevant figure and misplaces a relevant one. In contrast, VIG-RL-8B (Figure~\ref{fig:case1}) retrieves evidence from both modalities and accurately anchors verified figures to the corresponding content, demonstrating superior multimodal grounding.

\section{Conclusion}

In this paper, we tackle Verified Image Grounding (VIG)—the precise integration of retrieved visual evidence into generated text streams. Overcoming static baselines that fail to dynamically decide when to search and where to insert images, we propose VIG-RL. This autonomous agentic framework formulates the search-selection-insertion workflow as an active decision-making process. Optimized via GRPO within a ReAct-style loop, VIG-RL pairs intermediate retrieval supervision with strictly decoupled text-image terminal evaluation, achieving state-of-the-art performance and internalizing a robust search-and-insert policy.

\bibliography{aaai2027}
\clearpage

\section{Appendix}
\subsection{Implementation Details.}
\label{train}
We employ BGE-M3~\citep{bge-m3} as the unified retrieval backbone for both textual and visual retrieval. For the latter, rather than computing direct cross-modal similarity (i.e., CLIP-style text-to-image matching), we introduce \textit{Context-Anchored Image Retrieval} to preserve semantic coupling in knowledge-intensive domains. Formally, we represent the multimodal corpus as a set of context-image pairs $\mathcal{C} = \{(t_j, i_j)\}$, where each image $i_j$ is strictly anchored to its surrounding textual context $t_j$. Given an image search query $q_{\text{img}}$, the environment fetches the top-$K$ visual assets by ranking these pairs based on the text-to-text cosine similarity between the query and the anchor texts:

\begin{small}$$\mathcal{I}_{\text{retrieved}} = \left\{ i \;\middle|\; (t, i) \in \mathop{\text{Top-}K}_{(t_j, i_j) \in \mathcal{C}} \cos\big(\mathcal{E}(q_{\text{img}}), \mathcal{E}(t_j)\big) \right\},$$\end{small}
where $\mathcal{E}(\cdot)$ denotes the BGE-M3 text encoder. This mechanism explicitly bypasses the semantic mismatch of direct visual embeddings (ablated later). 

Our RL training framework is implemented with veRL~\citep{sheng2024hybridflow,wu2025mmsearch} using GRPO. VIG-RL is trained for $20$ epochs ($720$ optimization steps) with a learning rate of $2\times10^{-6}$, \texttt{rollout.n}=8, and a KL penalty coefficient of $0$. The agent is allowed to perform at most $T_{\max}=5$ reasoning-action interaction rounds. 
%During retrieval, the text retriever returns the top-2 textual documents, while the image retriever returns the top-5 candidate images for each query.
The training set comprises $1.1$k curated samples, derived from an $8:2$ train-test split of the \texttt{Web}, \texttt{Wiki}, and \texttt{Arxiv} subsets of MRAMG-Bench~\citep{yu2025mramg}, inspected for cross-modal consistency.

To assess whether the gains of VIG-RL can be explained merely
by supervised exposure to agentic rollouts, we construct an SFT
baseline using the same Qwen3-VL-8B-Instruct initialization and
the same $1.1$k training instances as VIG-RL. Specifically, for
each training instance, we sample five rollouts from
Gemini-2.5-Flash under the same agentic workflow, action space,
retrieval environment, and maximum interaction horizon. We
compute the final-answer Image F1 score for each rollout and
retain the rollout with the highest score. We use Image F1 as an
objective, rule-based selection criterion that does not rely on
an LLM judge. This best-of-five rollout selection produces
exactly one trajectory per training instance, resulting in
$1.1$k SFT training trajectories.

We fine-tune the base model on the selected trajectories using
the LLaMA-Factory~\citep{zheng2024llamafactory} framework.
The model is trained for $3$ epochs with an effective batch size
of $32$ and a learning rate of $1\times10^{-5}$. We use a
cosine learning-rate scheduler with a warmup ratio of $0.03$
and apply a weight decay of $0.01$. The SFT and RL settings
therefore share the base-model initialization, training
instances, action space, and retrieval environment, while
contrasting offline imitation of best-of-five Gemini-generated
rollouts with on-policy reward-driven optimization.

\subsection{Datasets and Evaluation Metrics}
\label{data}

\paragraph{Datasets.}
We conduct experiments on MRAMG-Bench~\citep{yu2025mramg}.
MRAMG-Bench consists of six datasets covering diverse text-image resources and information-seeking scenarios.
Each example provides a user query paired with a ground-truth interleaved text-image response.
This structure allows us to evaluate whether a model can retrieve, select, and insert verified visual evidence into long-form multimodal responses.

The benchmark consists of six datasets:
\begin{itemize}
\item \textbf{\texttt{Web}} contains dual-entity web queries paired with two corresponding verified images, requiring multi-image grounding within a single response.

\item \textbf{\texttt{Wiki}} contains Wikipedia-style entity-centric examples that mainly require single-image factual grounding.

\item \textbf{\texttt{Wit}} provides broad-domain image-text examples with single-image grounding derived from web and Wikipedia resources.

\item \textbf{\texttt{Arxiv}} is collected from academic papers, where figures and diagrams are closely associated with technical textual contexts.

\item \textbf{\texttt{Manual}} contains instruction manuals with dense text-image content and diverse visual evidence types.

\item \textbf{\texttt{Recipe}} contains recipe-oriented multimodal examples, often requiring grounding over procedural text and visual cooking evidence.
\end{itemize}

Following our experimental setting, we use \texttt{Web}, \texttt{Wiki}, and \texttt{Arxiv} as in-domain datasets, and evaluate out-of-domain generalization on \texttt{Wit}, \texttt{Manual}, and \texttt{Recipe}.
This split allows us to test both the agent's ability to learn from familiar retrieval patterns and its robustness to different document structures, image densities, and multimodal information needs.

\paragraph{Evaluation Metrics.}
We evaluate models using both image-grounding metrics and overall response-quality metrics. 
\begin{itemize}

\item \textbf{F1} \citep{yu2025mramg} measures the image-level F1 score between the inserted image set $\mathcal{I}_{\text{ans}}$ and the ground-truth verified image set $\mathcal{I}^*$.
We first compute image-level Precision and Recall as:
\[
\text{Precision} =
\frac{|\mathcal{I}^* \cap \mathcal{I}_{\text{ans}}|}
{|\mathcal{I}_{\text{ans}}|}.
\]

\[
\text{Recall} =
\frac{|\mathcal{I}^* \cap \mathcal{I}_{\text{ans}}|}
{|\mathcal{I}^*|}.
\]

The final F1 score is then computed as:
\[
\text{F1} =
\frac{2 \times \text{Precision} \times \text{Recall}}
{\text{Precision} + \text{Recall}}.
\]
This metric evaluates whether the model selects the correct verified images for the final response.

\item \textbf{Comprehensive Score (C.S.)} evaluates the overall quality of the final interleaved multimodal response, jointly considering textual correctness, image relevance, insertion-position appropriateness, text--image alignment, and overall multimodal coherence. Following \citet{yu2025mramg}, we use GPT-4o as the LMM-based judge with the evaluation prompt shown in Table~\ref{answer_eval_prompt}. 

\item \textbf{Image Ordering Score}~\citep{yu2025mramg} is additionally
reported for the \texttt{Manual} and \texttt{Recipe} datasets, whose
responses often contain multiple images with a meaningful procedural
order. Based on the weighted edit distance between the generated and
ground-truth image sequences, this metric assesses whether the selected
images follow the correct relative or procedural order, complementing
the broader assessment of image placement and text--image coherence
provided by C.S.

Let the ground-truth and generated image sequences be defined as follows:
\begin{description}
    \item[\textbf{Ground truth:}]
    $\mathcal{I}^{*}
    = i_1^{*} \rightarrow i_2^{*} \rightarrow \cdots
    \rightarrow i_n^{*}$,
    where $i_j^{*}$ denotes the image at the $j$-th position in the
    ground-truth sequence.

    \item[\textbf{Prediction:}]
    $\mathcal{I}
    = i_1 \rightarrow i_2 \rightarrow \cdots \rightarrow i_m$,
    where a predicted image $i_j$ is not necessarily contained in
    $\mathcal{I}^{*}$, and the number of predicted images $m$ may
    differ from the number of ground-truth images $n$.
\end{description}

The Order Score is computed as follows:
\begin{equation}
\begin{aligned}
\text{Order Score}
&=
\frac{|\mathcal{I}^{*} \cap \mathcal{I}|}{n}
\\
&\quad\times
\left(
1-\frac{1}{p}
\min\left(
\frac{
    \operatorname{dist}(\mathcal{I}^{*},\mathcal{I})
}{
    \max(n,m)
},
p
\right)
\right).
\end{aligned}
\end{equation}

The first term measures the coverage of ground-truth images in the
generated response, while the second term penalizes discrepancies
between the generated and ground-truth image orders.

Here, $\operatorname{dist}(\mathcal{S},\mathcal{S}')$ denotes the
weighted edit distance between two image sequences. It is defined as
the minimum total cost required to transform $\mathcal{S}'$ into
$\mathcal{S}$ using the following operations:
\begin{description}
    \item[\textbf{Insertion:}]
    Insert into $\mathcal{S}'$ an image that appears in $\mathcal{S}$
    but is missing from $\mathcal{S}'$. The operation cost is $p_1$.

    \item[\textbf{Deletion:}]
    Delete from $\mathcal{S}'$ an image that does not appear in
    $\mathcal{S}$. The operation cost is $p_2$.

    \item[\textbf{Substitution:}]
    Replace an image in $\mathcal{S}'$ with the corresponding image
    from $\mathcal{S}$ at the appropriate position. The operation
    cost is $p_3$.
\end{description}

The operation costs satisfy $p_1 > p_2 > p_3$. The normalization
constant $p \geq p_1$ ensures that the final Order Score falls within
the range $[0,1]$. The weighted edit distance can be computed using
dynamic programming with a time complexity of $O(mn)$.

\end{itemize}

Unless otherwise specified, all metrics in tables and figures are reported on a 0–100 scale.
\newpage

\subsection{Prompt Details}
\label{prompt}
The LLM prompts used in this study are presented below.

\begin{table*}[t]
\begin{center}
\colorbox{c7}{
\parbox{0.95\textwidth}{
\textbf{\#Task}

Imagine that you are a multimodal large language model proficient in
processing text--image inputs and producing interleaved text--image outputs.
You will be given a question and must generate a comprehensive multimodal
response iteratively.

\textbf{\#Action Space}

At each iteration, your action should include two parts:
First, you must conduct reasoning inside
\textcolor{c3}{<thinking>}...\textcolor{c3}{</thinking>}.
After reasoning, choose exactly one of the following three options as the
second part:

1. If you determine that visual evidence would improve the accuracy or clarity
of the answer, you can invoke an image search engine using
\textcolor{c1}{<img\_search>}query\textcolor{c1}{</img\_search>}.
It will return the top-ranked retrieved images and their associated information
between
\textcolor{c5}{<retrieved\_img>} and
\textcolor{c5}{</retrieved\_img>}.

2. If additional factual or contextual information is needed, you can invoke
a text search engine using
\textcolor{c2}{<text\_search>}query\textcolor{c2}{</text\_search>}.
It will return the top-ranked textual results between
\textcolor{c8}{<retrieved\_text>} and
\textcolor{c8}{</retrieved\_text>}.

3. If no further external information or visual evidence is required, provide
the final response inside
\textcolor{b1}{<answer>}...\textcolor{b1}{</answer>}
without any additional explanation. Assess the relevance of the retrieved
images and select suitable ones. While generating the textual response,
determine the most appropriate placement for each selected image. Within
\textcolor{b1}{<answer>}, integrate the selected images naturally into the
narrative by placing <img:id> at the most appropriate positions. Use only
image IDs that appear in
\textcolor{c5}{<retrieved\_img>}; never invent image IDs.

\textbf{\#Action Examples}

\textbf{\#\# Example 1}

\textcolor{c3}{<thinking>}your reasoning process\textcolor{c3}{</thinking>}
\textcolor{c1}{<img\_search>}key differences between mitosis and
meiosis\textcolor{c1}{</img\_search>}

\textbf{\#\# Example 2}

\textcolor{c3}{<thinking>}your reasoning process\textcolor{c3}{</thinking>}
\textcolor{c2}{<text\_search>}recommended daily protein intake in grams per
kilogram\textcolor{c2}{</text\_search>}

\textbf{\#\# Example 3}

\textcolor{c3}{<thinking>}your reasoning process\textcolor{c3}{</thinking>}
\textcolor{b1}{<answer>}Doing household chores helps maintain a clean and
comfortable home. In the kitchen, the dishes have been washed and placed
neatly in the drying rack, ready to be put away once they are dry.<img:152>
Similarly, in the living room, the sofa cushions have been fluffed and
arranged neatly, creating a comfortable space for relaxation.<img:100>
\textcolor{b1}{</answer>}\\

\textbf{\#Input}

Question: \{query\}
}
}

\end{center}
\caption{Prompt template used for training the VIG-RL model.}
\label{train_prompt}
\end{table*}

\begin{table*}[t]
\begin{center}
\colorbox{c7}{
\parbox{0.95\textwidth}{
\textbf{\#Task}

Imagine that you are a multimodal large language model proficient in
processing text--image inputs and producing interleaved text--image outputs.
You will be given a question and must generate a comprehensive multimodal
response iteratively.

\textbf{\#Action Space}

At each iteration, your action should include two parts:
First, you must conduct reasoning inside
\textcolor{c3}{<thinking>}...\textcolor{c3}{</thinking>}.
After reasoning, choose exactly one of the following two options as the
second part:

1. If you determine that visual evidence would improve the accuracy or clarity
of the answer, you can invoke an image search engine using
\textcolor{c1}{<img\_search>}query\textcolor{c1}{</img\_search>}.
It will return the top-ranked retrieved images and their associated information
between
\textcolor{c5}{<retrieved\_img>} and
\textcolor{c5}{</retrieved\_img>}.

2. If no further visual evidence is required, provide the final response inside
\textcolor{b1}{<answer>}...\textcolor{b1}{</answer>}
without any additional explanation. Assess the relevance of the retrieved
images and select suitable ones. While generating the textual response,
determine the most appropriate placement for each selected image. Within
\textcolor{b1}{<answer>}, integrate the selected images naturally into the
narrative by placing <img:id> at the most appropriate positions. Use only
image IDs that appear in
\textcolor{c5}{<retrieved\_img>}; never invent image IDs.

\textbf{\#Action Examples}

\textbf{\#\# Example 1}

\textcolor{c3}{<thinking>}your reasoning process\textcolor{c3}{</thinking>}
\textcolor{c1}{<img\_search>}key differences between mitosis and
meiosis\textcolor{c1}{</img\_search>}

\textbf{\#\# Example 2}

\textcolor{c3}{<thinking>}your reasoning process\textcolor{c3}{</thinking>}
\textcolor{b1}{<answer>}Doing household chores helps maintain a clean and
comfortable home. In the kitchen, the dishes have been washed and placed
neatly in the drying rack, ready to be put away once they are dry.<img:152>
Similarly, in the living room, the sofa cushions have been fluffed and
arranged neatly, creating a comfortable space for relaxation.<img:100>
\textcolor{b1}{</answer>}\\

\textbf{\#Input}

Question: \{query\}
}
}
\end{center}
\caption{Prompt template used for VIG-RL training without textual retrieval.}
\label{image_search_prompt}
\end{table*}

\begin{table*}[t]
\begin{center}
\colorbox{c7}{
\parbox{0.95\textwidth}{

\textbf{\#Task}
Imagine you are a multimodal large model proficient in processing text-image input and providing interwoven text-image responses. You will receive a context that includes several images represented as placeholders, along with a query related to the given context. Your task is to select appropriate images from the provided context (if none are suitable, you may choose not to include any) and generate a mixed media response to the query, combining text and the selected images. Please note, your answer should be presented in an interwoven text-image format, where you select images from the context and output them in the corresponding placeholder format. Please provide only the answer, without including any analysis.
Each candidate image is associated with a unique identifier in the format <img:id>.
Image Insert: When inserting image placeholders, place them at the most appropriate point within the answer. Image placeholders should be embedded naturally in the answer to support and enhance understanding, such as when describing specific locations, historical events, or notable buildings. Use only image placeholders that explicitly appear in the provided context or image-caption list. Never invent, modify, or refer to an image placeholder that is not provided.\\
\textbf{\# Output Format}
Please output the answer in an interwoven text-image format, where you select images from the context provided and output them in the corresponding placeholder format.\\
\textbf{\# Output Example}
Doing household chores is a daily task that helps maintain a clean home. At the kitchen, dishes are neatly washed and placed in the drying rack, ready to be put away once they dry.<img:152> Similarly, in the living room, the sofa cushions are fluffed and arranged properly, creating a comfortable space for relaxation.<img:100>

\textbf{\#Input}

Question:\{query\}\\
Context:\{context\}\\
Image Caption:\{image\_captions\}
}
}
\end{center}
\caption{Answer Generation Prompt for RAG workflow.}
\label{rag_prompt}
\end{table*}

\begin{table*}[t]
\begin{center}
\colorbox{c7}{
\parbox{0.95\textwidth}{

Your job is to act as an Evaluation Judge. You must look at a question, a gold textual answer, and a predicted textual answer, and assign a score from 1 to 5, where 5 is the highest.\\

\textbf{Scoring Definitions (1-5 Scale)}\\
Score 5 - Excellent (Fully Correct): The predicted answer is fully accurate, complete, and directly addresses the question. It contains all essential information present in the gold target and introduces NO factual errors or contradictions. Differences in wording, order, or phrasing do not matter.\\
Score 4 - Mostly Correct (Minor Omission/Flaw): The predicted answer is largely correct. It may have a very minor omission of non-critical detail OR a small phrasing issue that does not impact correctness. The main information is correct and there are NO factual errors or contradictions.\\
Score 3 - Partially Correct (Attempted): The predicted answer provides some correct and relevant information, but is significantly incomplete, vague, OR missing critical parts of the required answer. It contains NO factual errors or contradictions, but is insufficient for a higher score.\\
Score 2 - Mostly Incorrect (Significant Error): The predicted answer attempts to address the question but includes significant factual errors, contradictions, or misleading information. Some correct elements may be present, but the response is overall unreliable.\\
Score 1 - Failure (Contradiction or Not Attempted): The predicted answer either directly contradicts the gold target or known facts, OR is irrelevant, nonsensical, empty, or otherwise fails to attempt the question.\\

\textbf{Additional Grading Rules}\\
Scope: Only evaluate the predicted answer based on what the question asks. If the gold target includes extra, unrequested detail, the prediction is not penalized for omitting it.
Numeric Answers: Numeric correctness is determined by matching the gold target's value and significant figures. Minor deviations or incorrect significant figures typically result in Score 2.
Typos: Minor typos or naming variations are acceptable if the meaning is clear.
Mixed Information: If the response contains both correct and significantly incorrect information, assign Score 2.\\
\textbf{\#Evaluation Input}\\
Question: \{question\} \\
Gold textual answer: \{correct\_answer\_text\}\\
Predicted textual answer: \{response\_text\}\\
Output only the single integer score (1, 2, 3, 4, or 5). No explanation or other text.
}
}
\end{center}
\caption{Prompt template used for answer evaluation during training.}
\label{train_eval_prompt}
\end{table*}

\begin{table*}[t]
\begin{center}
\colorbox{c7}{
\parbox{0.95\textwidth}{

Your job is to act as an Evaluation Judge. You must look at a question, a gold target, and a predicted answer, and assign a score from 1 to 5, where 5 is the highest. The gold target and predicted answer may contain interleaved text and images. You must evaluate both the textual content and the actual visual content of the images.

\textbf{Scoring Definitions (1-5 Scale)}\\
Score 5 - Excellent (Fully Correct): The predicted answer is fully accurate, complete, and directly addresses the question. Its text is correct, all necessary images are correct and relevant, and the images are appropriately placed and aligned with the surrounding text. It contains all essential information present in the gold target and introduces NO factual errors, visual errors, or contradictions. Differences in wording, order, or phrasing do not matter unless the order is semantically or procedurally important.

Score 4 - Mostly Correct (Minor Omission/Flaw): The predicted answer is largely correct. It may have a very minor omission of non-critical textual or visual detail, OR a small phrasing, image-placement, or text-image alignment issue that does not substantially impact correctness. The main information and visual evidence are correct, and there are NO significant factual errors or contradictions.

Score 3 - Partially Correct (Attempted): The predicted answer provides some correct and relevant textual or visual information, but is significantly incomplete, vague, missing critical information or images, OR contains noticeable image-placement or text-image alignment problems. It is insufficient for a higher score but still provides meaningful correct information.

Score 2 - Mostly Incorrect (Significant Error): The predicted answer attempts to address the question but includes significant factual or visual errors, contradictions, misleading information, incorrect or irrelevant images, or serious text-image misalignment. Some correct elements may be present, but the response is overall unreliable.

Score 1 - Failure (Contradiction or Not Attempted): The predicted answer either directly contradicts the gold target or known facts, uses images that are largely unrelated or misleading, OR is irrelevant, nonsensical, empty, or otherwise fails to attempt the question.

\textbf{Additional Grading Rules}\\
Scope: Only evaluate the predicted answer based on what the question asks. If the gold target includes extra, unrequested detail, the prediction is not penalized for omitting it.

Image Evaluation: Evaluate whether the predicted images are correct and relevant, whether necessary images are missing, whether each image is placed at an appropriate semantic position, and whether it supports the surrounding text. For procedural or ordered content, also evaluate whether the relative image order is correct. Do not require an exact token-level insertion-position match.

Visual Content: Evaluate the actual visible content of the images. Do not determine visual correctness solely from image identifiers or placeholder names.

Alternative Evidence: Semantically equivalent images may be accepted if they provide equally correct and relevant visual evidence, unless the question explicitly requires the exact gold image.

No-Image Cases: If neither answer contains images and the question does not require visual evidence, evaluate only the textual content. If necessary visual evidence is missing from the predicted answer, lower the score accordingly.

Numeric Answers: Numeric correctness is determined by matching the gold target's value and required precision. Reasonable rounding is acceptable unless the question explicitly requires exact significant figures.

Typos: Minor typos or naming variations are acceptable if the meaning is clear.

Mixed Information: If the response contains both correct and significantly incorrect textual or visual information, assign Score 2.

\textbf{\#Evaluation Input}\\
Question: \{question\}\\
Gold target: \{correct\_answer\}\\
Predicted answer: \{response\}\\
Image captions: \{image\_captions\}\\
The images are provided in the following order: first, all images appearing in the Gold target, followed by all images appearing in the Predicted answer. Within each answer, the images follow the order of their image placeholders. The image captions correspond one-to-one with the images and are provided in the same order.
Output only the single integer score (1, 2, 3, 4, or 5). No explanation or other text.
}
} 
\end{center}
\caption{Prompt template used for holistic multimodal answer evaluation in the MLLM-reward ablation.
}
\label{train_eval_image_prompt}
\end{table*}

\begin{table*}[t]
\begin{center}
\colorbox{c7}{
\parbox{0.95\textwidth}{
You are an evaluator for interleaved multimodal responses.\\

You will be given:\\
- a user Question.\\
- a Gold Answer.\\
- and a Predicted Answer.\\
- the actual images appearing in the Gold Answer and Predicted Answer.\\
- and the corresponding image captions.\\

The images are ordered as follows: all images from the Gold Answer first,
followed by all images from the Predicted Answer. Within each answer,
the images follow the order of their placeholders. The captions correspond
one-to-one with the images and are provided in the same order. Each image
is referenced by a unique placeholder such as <img:id>.

Your task is to evaluate the overall quality of the Predicted Answer.\\

Image Evaluation:\\
- Evaluate the actual visual content of the images used in the Predicted Answer, as well as the information conveyed by those images.\\
- Use image IDs to associate images across the Gold and Predicted Answers. For image selection, compare the deduplicated ID sets, but do not judge visual correctness solely from the identifiers.\\
- Evaluate image placement by checking whether each selected image appears at an appropriate semantic position in the response.\\
- Evaluate text-image alignment by checking whether each image supports or illustrates the surrounding textual content.\\
- Evaluate image ordering by comparing the relative order of the selected images with the logical or procedural order in the Gold Answer.\\
- Do not require an exact token-level position match;  evaluate whether the semantic placement is appropriate.\\
- Duplicate occurrences of the same image should not affect image-selection matching, but unnecessary repetition should be penalized under placement and presentation quality.\\
- Any referenced image ID that is not present in the provided image list should be treated as an incorrect image.\\
- Strongly reward correct image selection, appropriate placement, and coherent text-image alignment.\\
- Strongly penalize incorrect, missing, misplaced, or incorrectly ordered images.\\

Textual Consistency:\\
- Check whether the Predicted Answer preserves the main meaning of the Gold Answer.\\
- Penalize semantic inconsistency or missing key information.\\
- Minor wording differences are acceptable.\\
- Good text alone should NOT receive a high score if the images are wrong.\\

Scoring Reference (0-1):\\

- 1.0: The textual response is fully correct; all necessary visual evidence is correct and relevant; and the images are appropriately placed, ordered, and aligned with the surrounding text.\\
- 0.8:
  Image selection is nearly perfect with only minor omissions,
  and the text is mostly correct.\\
- 0.5:
  Some images are correct but important images are missing or partially incorrect.
  Text may still be reasonably consistent.\\
- 0.2:
  Many important images are incorrect or missing,
  even if parts of the text are correct.\\
- 0:
  Images are completely wrong, severely mismatched,
  or largely unrelated to the Gold Answer.
  Use this range even if the text appears reasonable.\\

Special Rule:
- If both Gold Answer and Predicted Answer contain no images,
  evaluate only textual consistency.
\\

\textbf{Evaluation Input}\\

Question: \{question\}\\
Gold Answer: \{correct\_answer\}\\
Predicted Answer: \{response\}\\
Image captions: \{image\_captions\}\\
Output only the final score as a float between 0 and 1 rounded to 3 decimal places.
Do not output explanations or additional text.
}
}
\end{center}
\caption{Prompt template used to evaluate the Comprehensive Score of interleaved multimodal responses.}
\label{answer_eval_prompt}
\end{table*}

\subsection{Case Analysis}
\label{case_show}

In this section, we provide representative examples to illustrate the differences in output quality between VIG-RL and existing baselines.

\begin{figure*}[t]
\centering
\includegraphics[width=0.7\linewidth]{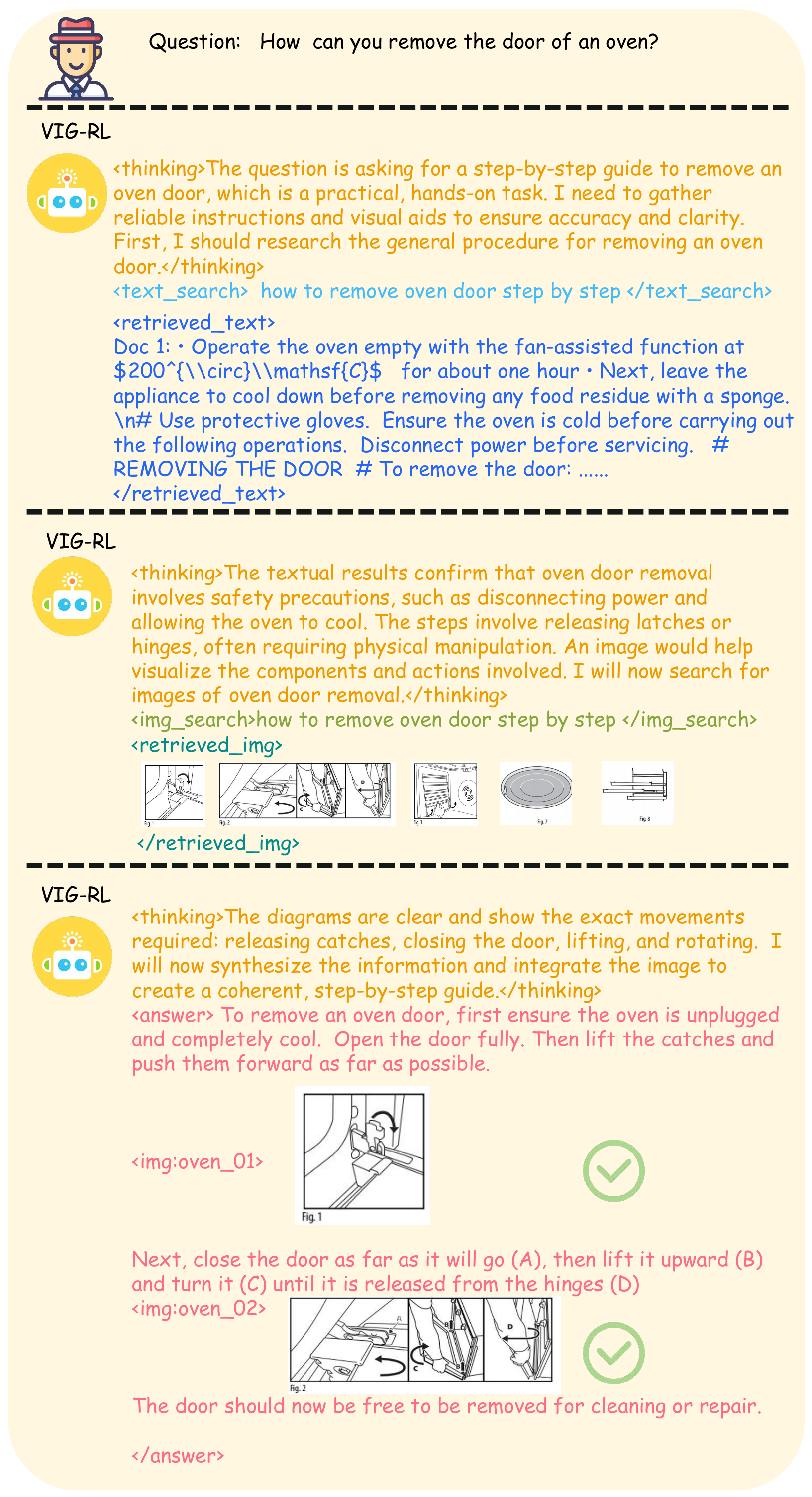}
\caption{
Case study on the \texttt{Manual} dataset. VIG-RL-8B retrieves both textual evidence and verified figures from the source manual and inserts the figures at the corresponding procedural steps, demonstrating its ability to jointly perform evidence retrieval, image selection, and text-image alignment.
}
\label{fig:case1}
\end{figure*}

As shown in Figure~\ref{fig:case1}, VIG-RL retrieves the relevant manual instructions and the two verified figures required for the door-removal procedure. It places each figure at the corresponding procedural step: the first supports lifting and pushing the catches forward, while the second illustrates closing, lifting, turning, and releasing the door. This shows that VIG-RL aligns retrieved visual evidence with the local semantics of the answer, rather than simply appending images.

\begin{figure*}[t]
\centering
\includegraphics[width=0.7\linewidth]{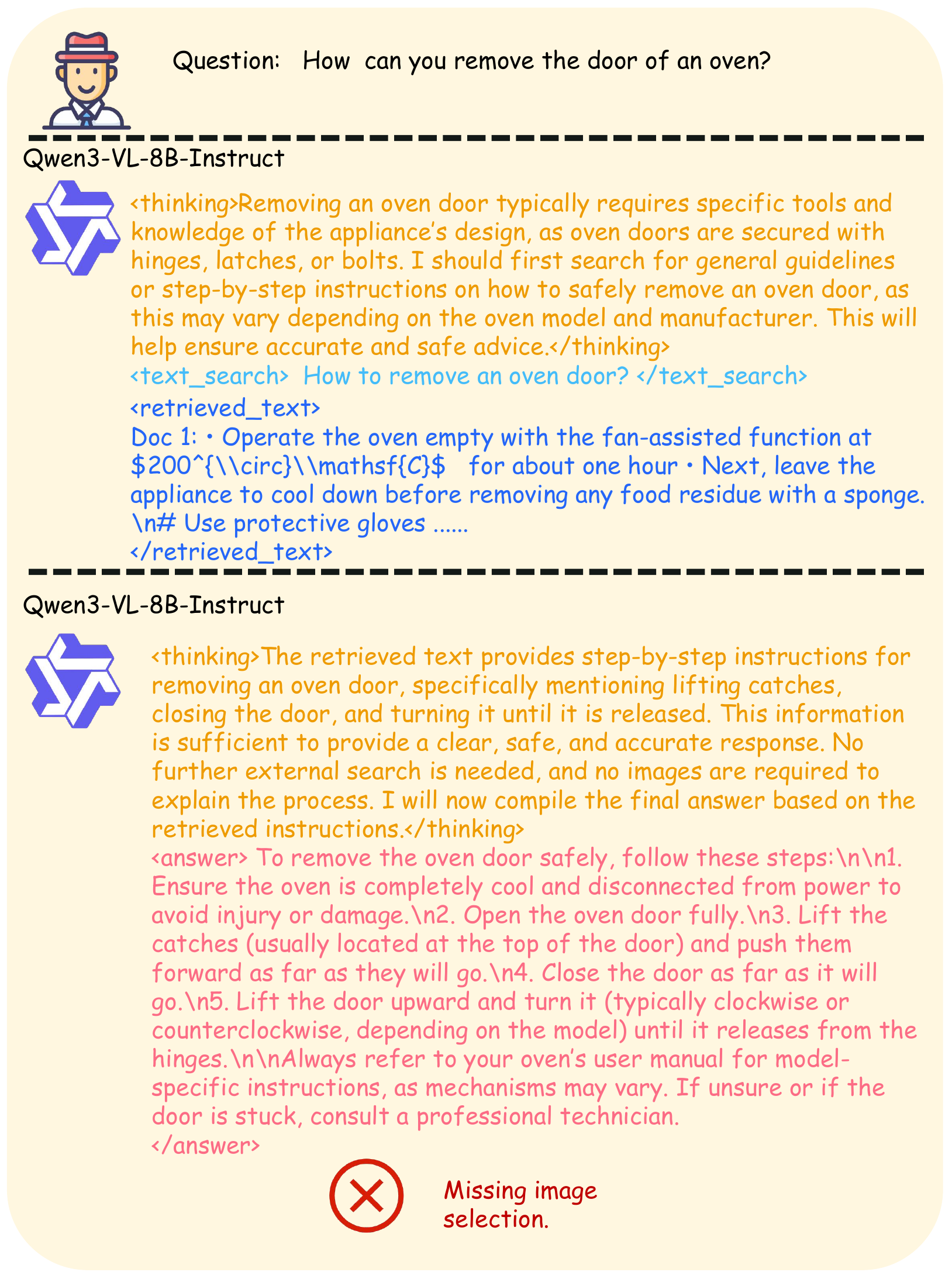}
\caption{
Case of Qwen3-VL-8B-Instruct under the agentic setting on the \texttt{Manual} dataset. Although the model retrieves partially relevant textual evidence from the source manual, it fails to further retrieve and select the verified figures required by the procedure, producing a text-only answer with missing visual grounding.
}
\label{fig:case2}
\end{figure*}

\begin{figure*}[t]
\centering
\includegraphics[width=0.7\linewidth]{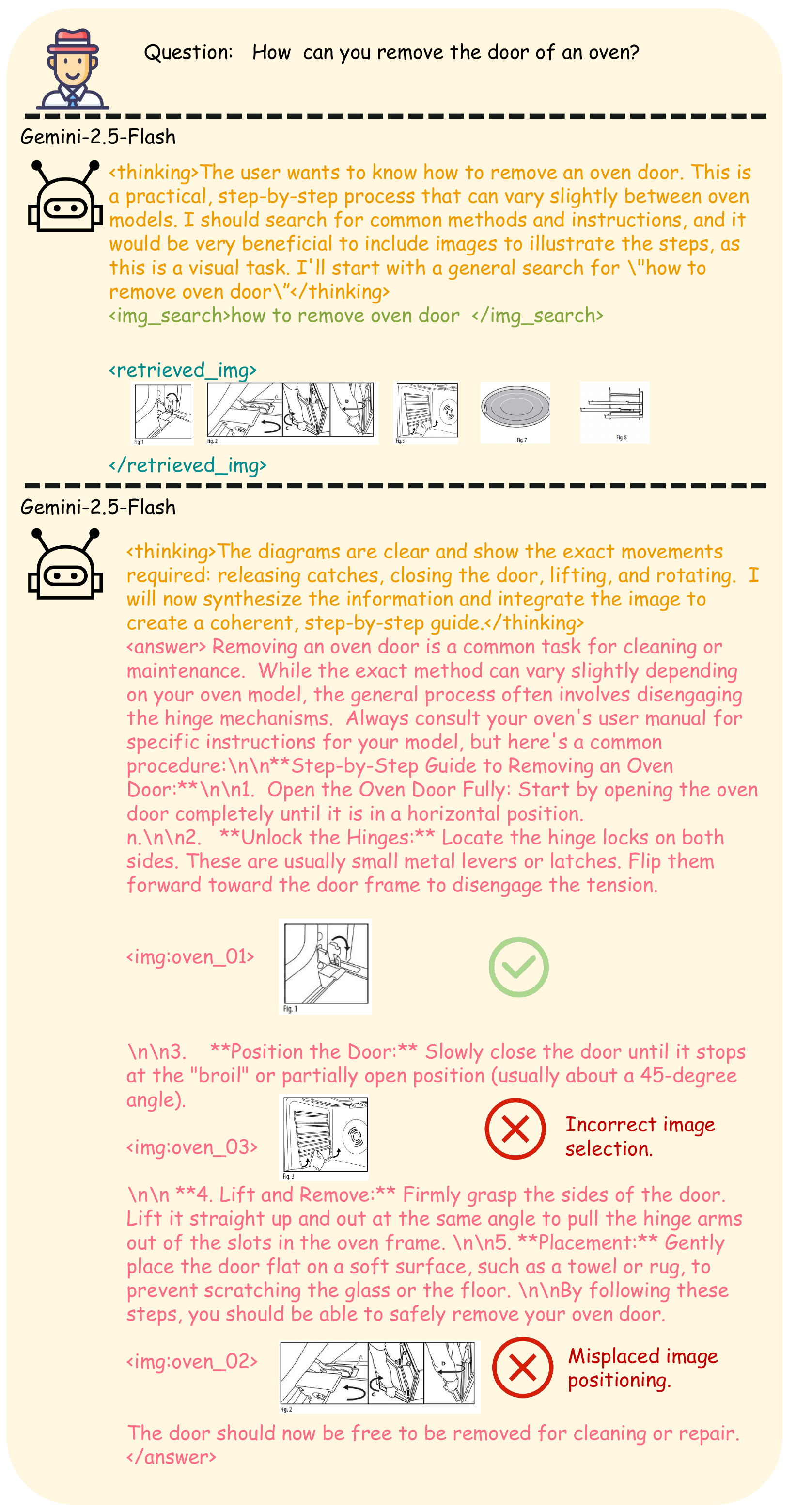}
\caption{
Case of Gemini-2.5-Flash under the agentic setting on the \texttt{Manual} dataset. Although the model performs image search and inserts retrieved figures into the response, it selects an irrelevant figure and places a relevant figure at an incorrect procedural position, leading to inaccurate visual grounding.
}
\label{fig:case3}
\end{figure*}

Figures~\ref{fig:case2} and~\ref{fig:case3} illustrate two representative failure modes of existing agentic baselines. Qwen3-VL-8B-Instruct retrieves partially relevant textual evidence but stops after text search, producing a text-only answer without the required verified figures. Gemini-2.5-Flash performs image search and inserts images, but selects an irrelevant figure and places a relevant one at an incorrect step. These cases show that tool use alone is insufficient for verified image grounding: the model must learn not only when to search, but also how to select and align visual evidence.

Overall, this example highlights the benefit of learning the complete search-and-insert behavior. Compared with general agentic models, VIG-RL more effectively retrieves textual evidence, selects verified figures, and integrates them into semantically appropriate positions, leading to more faithful and useful interleaved responses.

\end{document}